\documentclass[apj]{emulateapj}
\usepackage{longtable}
\usepackage{amsmath}

\begin{document}

\title{Gemini/GMOS Spectroscopy of 26 Strong Lensing Selected Galaxy Cluster Cores\altaffilmark{*}}
\author{
Matthew B. Bayliss\altaffilmark{1,2}, 
Joseph F. Hennawi\altaffilmark{3},
Michael D. Gladders\altaffilmark{1,2},
Benjamin P. Koester\altaffilmark{1,2}, 
Keren Sharon\altaffilmark{2}, 
H{\aa}kon Dahle\altaffilmark{4}, 
Masamune Oguri\altaffilmark{5}
}

\email{mbayliss@oddjob.uchicago.edu}

\altaffiltext{*}{Based on observations obtained at the Gemini
Observatory, which is operated by the Association of Universities
for Research in Astronomy, Inc., under a cooperative agreement
with the NSF on behalf of the Gemini partnership: The United States, 
The United Kingdom, Canada, Chile, Australia, Brazil and Argentina, 
with supporting data collected at the Subaru Telescope, operated
by the National Astronomical Observatory of Japan; the 2.5m Nordic   
Optical Telescope, operated on the island of La Palma jointly by
Denmark, Finland, Iceland, Norway, and Sweden, in the Spanish
Observatorio del Roque de los Muchachos of the Instituto de
Astrofisica de Canarias; the 3.5m Wisconsin-Indian-Yale-NOAO       
Telescope, at the WIYN Observatory which is a joint facility of the
University of Wisconsin-Madison, Indiana University, Yale University,
and the National Optical Astronomy Observatory; and the Apache Point 
Observatory 3.5-meter telescope, which is owned and operated by the 
Astrophysical Research Consortium.}
\altaffiltext{1}{Department of Astronomy \& Astrophysics, University of Chicago, 5640 South Ellis Avenue, Chicago, IL 60637}
\altaffiltext{2}{Kavli Institute for Cosmological Physics, University of Chicago, 5640 South Ellis Avenue, Chicago, IL 60637}
\altaffiltext{3}{Max-Planck-Institut f{\"u}r Astronomie K{\"o}nigstuhl 17, D-69117, Heidelberg, Germany}
\altaffiltext{4}{Institute of Theoretical Astrophysics, University of Oslo, P.O. Box 1029, Blindern, N-0315 Oslo, Norway}
\altaffiltext{5}{Division of Theoretical Astronomy, National Astronomical Observatory of Japan, 2-21-1 Osawa, Mitako, Tokyo 181-8588, Japan}

\begin{abstract}

We present results from a spectroscopic program targeting $26$ strong 
lensing cluster cores that were visually identified in the Sloan 
Digital Sky Survey (SDSS) and the Second Red-Sequence Cluster Survey 
(RCS-2). The $26$ galaxy cluster lenses span a redshift range of 
$0.2 < z < 0.65$, and our spectroscopy reveals $69$ unique background 
sources with redshifts as high as $z=5.200$. We also identify redshifts 
for $262$ cluster member galaxies and measure the velocity dispersions 
and dynamical masses for $18$ clusters where we have redshifts for 
$N\geq10$ cluster member galaxies. We include an accounting for the 
expected biases in dynamical masses of strong lensing selected clusters 
as predicted by results from numerical simulations and discuss possible 
sources of bias in our observations. The median dynamical mass of the $18$ 
clusters with $N\geq10$ spectroscopic cluster members is 
$M_{Vir} = 7.84\times10^{14} M_{\sun} h_{0.7}^{-1}$, which is somewhat 
higher than predictions for strong lensing selected clusters in simulations. 
The disagreement is not significant considering the large uncertainty in 
our dynamical data, systematic 
uncertainties in the velocity dispersion calibration, and limitations of 
the theoretical modeling. Nevertheless our study represents an important 
first step toward characterizing large samples of clusters that are 
identified in a systematic way as systems exhibiting dramatic strong 
lensing features.

\end{abstract}


\section{Introduction}

The evolution of large scale structure over cosmic time is a key test of 
the standard concordance cosmological model, and a tool for 
estimating cosmological parameters. Surveys designed to identify 
large samples of galaxy clusters are now producing catalogs of clusters 
with well-defined selection functions over large fractions of the sky 
\citep{bohringer2004,gladders2005,burenin2007,koester2007,vanderlinde2010}, 
and extensive efforts are underway to characterize observable proxies for cluster 
masses in order to convert cluster catalogs into robust measurements 
of cluster abundances as a function of mass and redshift 
\citep[e.g.,][]{vikhlinin2009a,vikhlinin2009b,rozo2009a,rozo2009b}.
Most observable quantities -- optical light, X-ray light, and the Sunyaev 
Zel'dovich (SZ) effect -- trace baryonic matter in clusters, but cluster 
mass and density profiles on large scales are dominated by dark matter. 
The dark matter content in galaxy clusters is most directly probed via the 
gravitational lensing effect; weak lensing measures the shape of the 
gravitational potential at relatively large radii while strong lensing 
provides detailed constraints on the mass structure within the cores 
of galaxy clusters. Weak lensing observations of galaxy clusters have 
become a powerful tool in recent years 
\citep{dahle2006,hoekstra2008,sheldon2009,okabe2009} but galaxy clusters 
exhibiting strong lensing remain are a rare subset of the larger population.

In this paper we present spectroscopic follow-up of a subset of a large sample
of several hundred giant arcs discovered in the SDSS \citep{york2000} and
RCS-2. Two forthcoming papers will describe the full giant arc samples
discovered in the SDSS (M.~D. Gladders et al. 2011, in prep) and the
RCS-2 (M.~B. Bayliss et al. 2011, in prep). These giant arc samples are intended
primarily to address the persistent lack of large, well-selected catalogs of
giant arcs which can be compared against $\Lambda$CDM predictions for giant
arc statistics, as well as to provide statistical samples of strong 
lensing clusters that can be used to study the detailed structure of 
cluster cores and mass distributions. A large sample of strong lensing 
clusters also increases the volume of the high-redshift universe that is 
available for observations with the aid of foreground cluster lenses 
serving as ``natural telescopes.'' From the data presented here we recover 
spectroscopic redshifts for a sample of $69$ background sources behind $26$ 
distinct cluster cores, many of which are obviously multiply imaged, and 
all of which are likely magnified by the foreground cluster potentials. These 
data represent a significant extension in the number of confirmed strong lensing 
clusters -- especially at $z\gtrsim0.2$ -- and provide a sample of cluster lenses 
that we use to test predictions for the characteristic masses of such systems.

Where necessary we assume a flat cosmology with $H_{0}=70$ km s$^{-1}$ 
Mpc$^{-1}$, $\sigma_{8}=0.81$ and matter density $\Omega_{M}=0.25$.

\section{Observations}

%number 1
\begin{figure}
\centering
\includegraphics[scale=0.67]{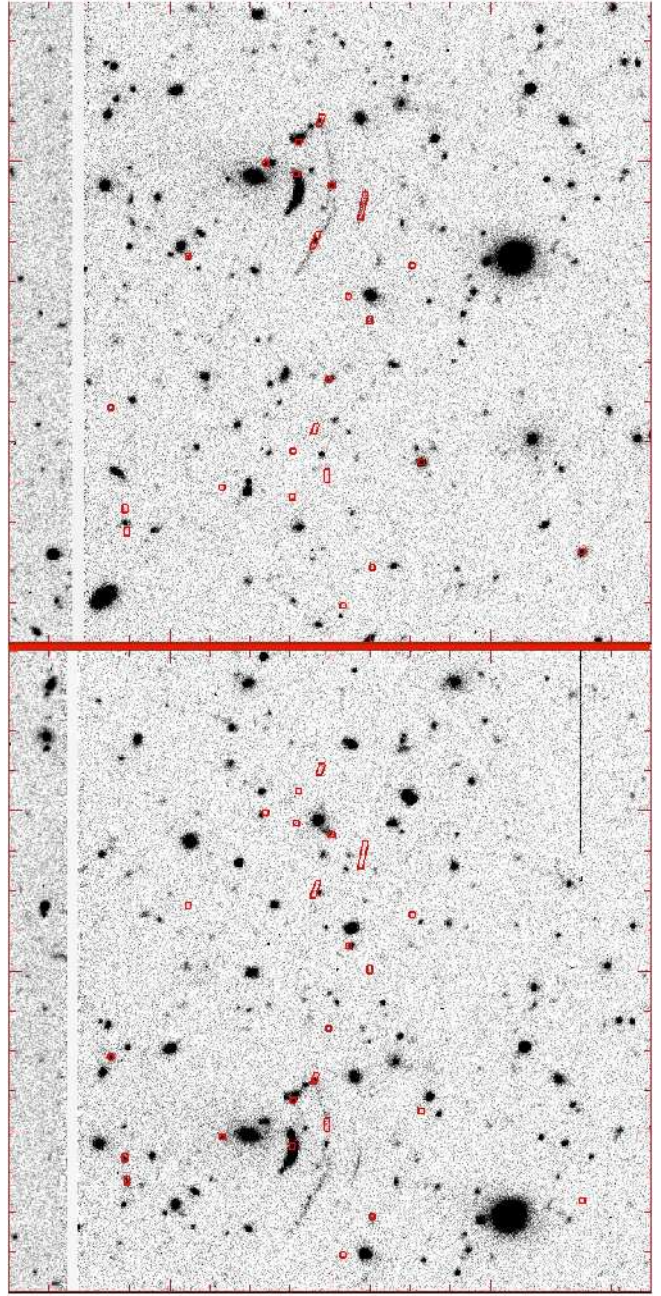}
\caption{\scriptsize{
Images of one of our strong lensing cluster targets, SDSS J1138+2754,
with the corresponding spectroscopic mask overlaid at each of the
pointing and nod positions. Note that some strong lensing features
are targeted with slits in both positions, ensuring that we collect
science data for those arcs during the entire N\&S exposure sequence.
We are also able to target multiple candidate strong lensing features
that would collide spectrally for a single standard multi-object
spectroscopic slitmask.
$Top:$ GMOS $r-band$ 300s image of the target cluster at the initial
pointing coordinates with slits overlaid in red.
$Bottom:$ GMOS $r-band$ 300s image of the target cluster at the nod
position with slits overlaid in red.}
}
\label{nodandshuffle}
\end{figure}

\subsection{Targeted Strong Lensing Clusters}

%number 2a
\begin{figure*}[t]
\centering
\includegraphics[scale=0.6]{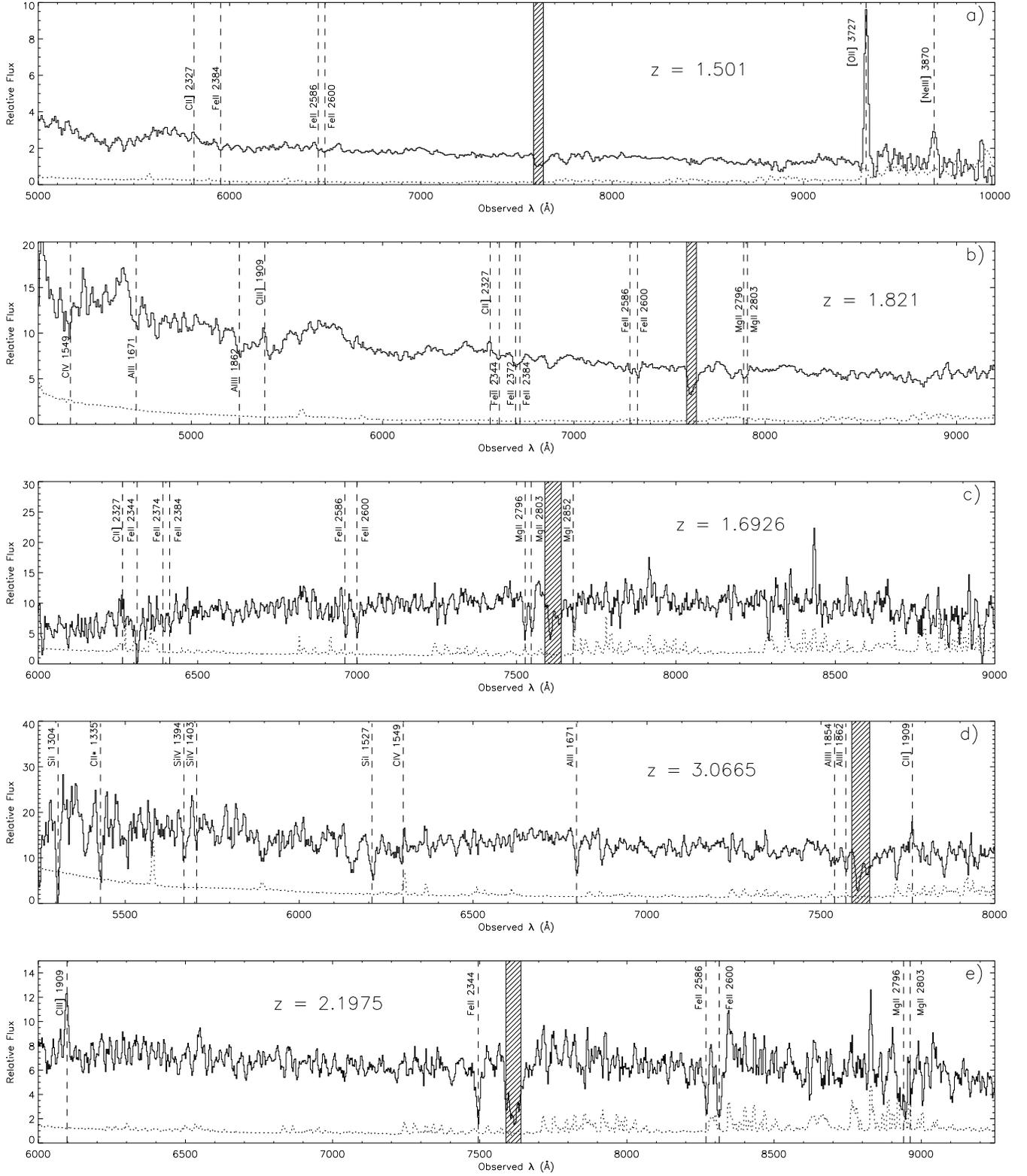}
\caption{\scriptsize{
Gemini/GMOS-North nod-and-shuffle spectra for five sources with
high confidence redshifts (class 3). Spectra
are displayed in the observer-frame, and smoothed to match the spectral
resolution of the data. The dotted histogram is the error array for the
spectra, and the locations of spectral lines are identified
by dashed lines and labeled with their corresponding ion and rest-frame
wavelength. The telluric A Band absorption feature is indicated by a
vertical shaded region. From top to bottom the spectra in each panel
correspond to the following sources in Table~\ref{arcproperties} -- $a)$
SDSS J0915+3826, object A2 in Figure \ref{color2}; $b)$ SDSS J0957+0509,
source A in Figure \ref{color7}; $c)$ SDSS J0851+3331, source A in
Figure \ref{color2}; $d)$ SDSS J1420+3955, source B in Figure \ref{color5};
$e)$ SDSS J1038+4849, source A in Figure \ref{color2}. Spectra in panels 
a and b are lower resolution data from our 2008A program, while spectra 
in panels c, d, and e are at higher spectral resolution and were taken as 
part of our 2009A program.}}
\label{gmosspec3}
\end{figure*}

%number 2b
\begin{figure*}[t]
\centering
\includegraphics[scale=0.6]{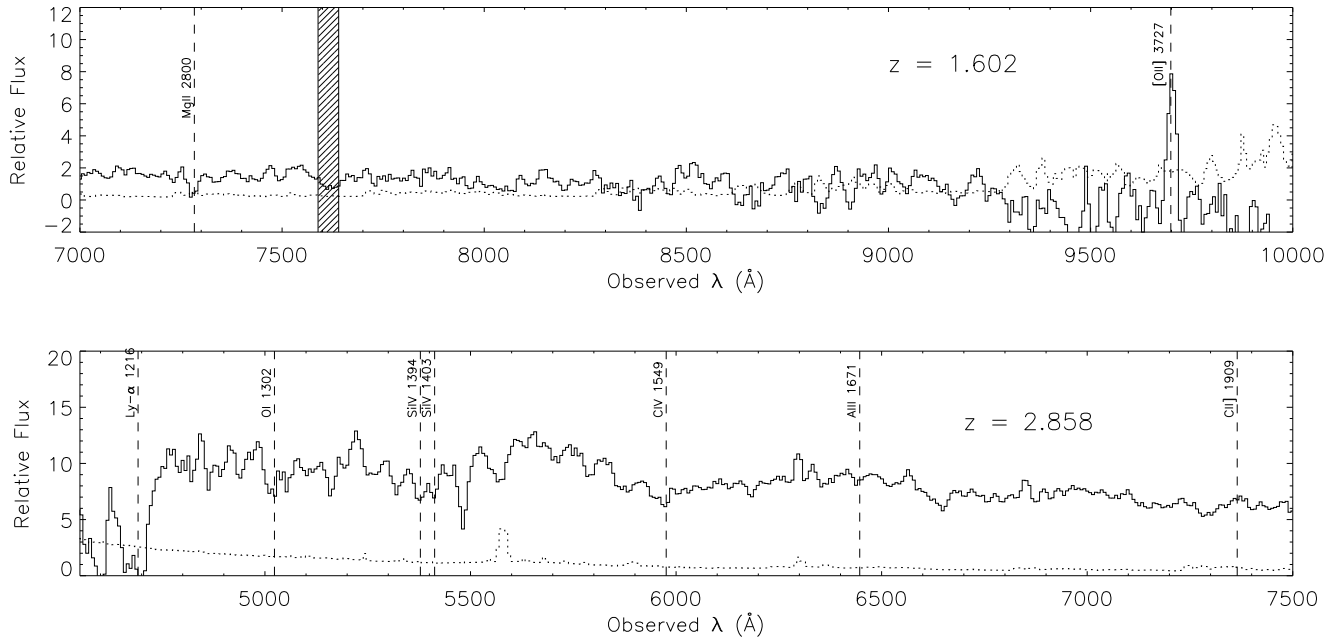}
\caption{\scriptsize{
Gemini/GMOS-North nod-and-shuffle spectra for two sources with
medium confidence redshifts (class 2). Spectra are
displayed in the same manner as in Figure ~\ref{gmosspec3}.
From top to bottom the spectra in each panel
correspond to the following sources in Table~\ref{tabredshifts} --
$Top:$ RXC J1327.0+0211, source C in Figure~\ref{color4},
$Bottom:$ SDSS J2111-0114, source A in Figure~\ref{color1}. 
Both spectra displayed here are from our 2008A program.
}}
\label{gmosspec2}
\end{figure*}

%number 2c
\begin{figure*}[t]
\centering
\includegraphics[scale=0.6]{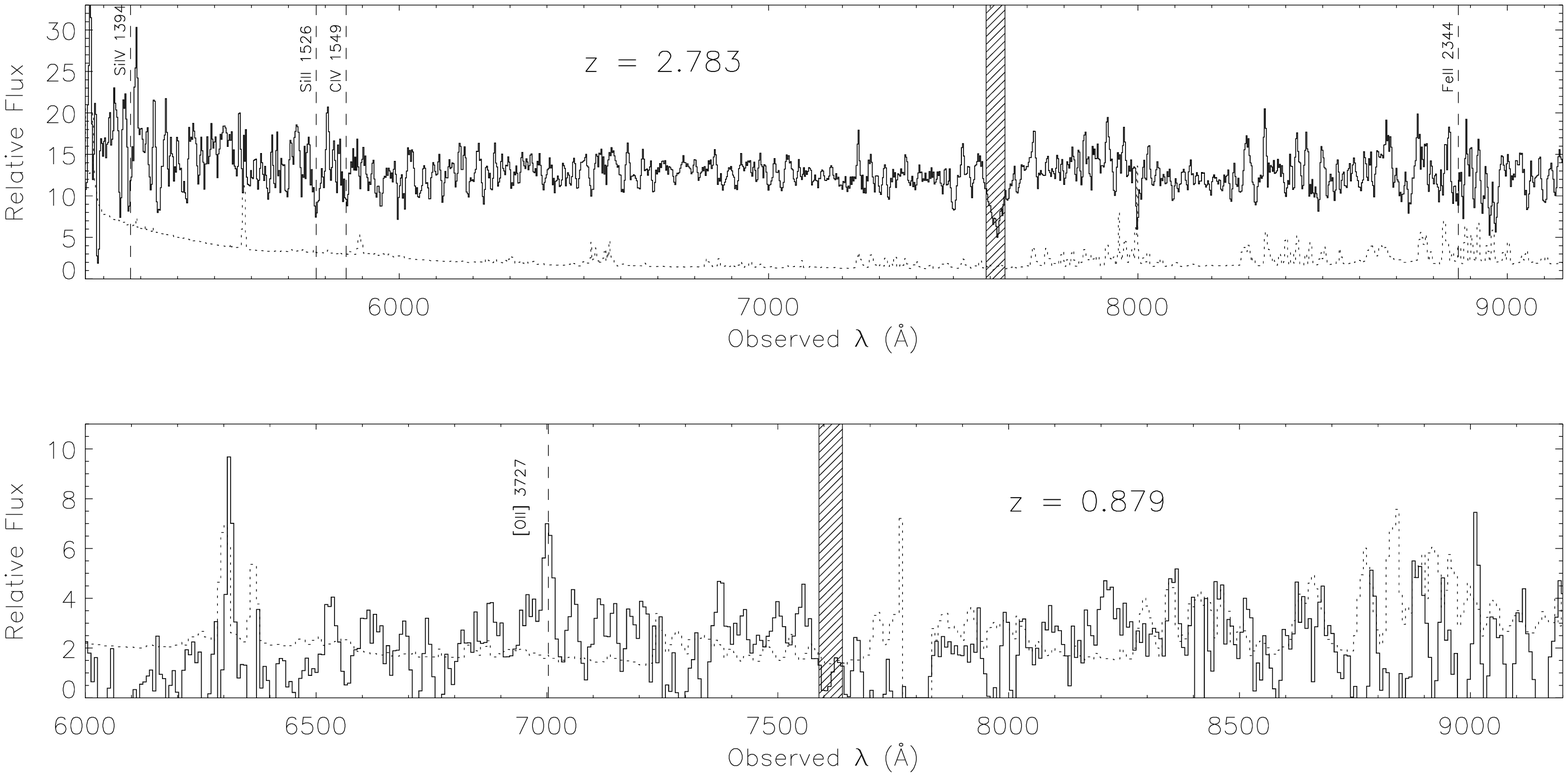}
\caption{\scriptsize{
Gemini/GMOS-North nod-and-shuffle spectra for two sources with
low confidence redshifts (class 1). Spectra are
displayed in the same manner as in Figure~\ref{gmosspec3}.
From top to bottom the spectra in each panel
correspond to the following sources in Table~\ref{tabredshifts} --
$Top:$ SDSS J1038+4849, source C in Figure~\ref{color2};
$Bottom:$ SDSS J1209+2640, source B in Figure~\ref{color4}. 
The spectrum in the top panel is from our 2009A program, and 
the spectrum on the bottom is from 2008A. The spectrum in the 
bottom panel is identified as $z=0.879$ by assuming the lone 
robust emission feature corresponds to [OII]~$\lambda3727$\AA, 
though we do not see corroborating [OIII]~$\lambda5007$\AA~ at 
9410\AA~, where the sky subtraction has large residuals. Sky line 
residuals were surprisingly large in our 2008A data in spite of 
the use of nod and shuffle, which was a strong motivator for the 
change in observational strategy between 2008A and 2009A.}}
\label{gmosspec1}
\end{figure*}

The targeted strong lensing clusters were initially identified in one of several 
visual searches for giant arcs in an exhaustive sample of red-sequence selected 
clusters in the SDSS and RCS-2 surveys. Our visual searches produced three 
distinct giant arc samples, each of which has different visual selection 
criteria. The SDSS ``Visual'' Sample (Gladders, M.~D. et al., in prep) is composed 
of candidate strong lensing clusters that were identified in the relatively 
shallow SDSS survey imaging. We confirmed the lensing interpretation in follow-up 
$g$-band imaging on $\sim2 - 4$m class telescopes. The SDSS ``Blind'' Sample 
consists of strong lensing clusters that were identified in follow-up $g-$band 
imaging of the most massive $\sim200$ clusters, as selected by the red sequence 
from the SDSS photometry \citep{hennawi2008}.  The RCS-2 Giant Arc sample 
(Bayliss, M.~B. et al., in prep) is defined in the same way as the SDSS Visual 
sample, but uses imaging data that is $\sim2$ magnitudes deeper than the SDSS, with 
a median seeing of $\sim0.7\arcsec$, and therefore facilitates morphological 
classifications on par with the follow-up imaging of the SDSS giant arcs. 
See \citet{gilbank2011} for a detailed description of the RCS-2 data.

We have adopted a naming convention for giant arcs discovered in the
SDSS-- Sloan Giant Arc Survey(SGAS) Jhhmmss+ddmmss \citep{koester2010} -- 
and giant arcs discovered in the RCS-2 -- Red-Sequence Cluster Survey Giant Arc 
(RCSGA) Jhhmmss+ddmmss \citep[e.g.,][]{wuyts2010}. These two surveys for giant 
arcs have produced hundreds of strong lensing clusters, and we followed-up a 
subset of these systems spectroscopically. We observed a sample of $26$ clusters 
with the Frederick C. Gillett Telescope (Gemini North) between February 2008 and 
June 2010 as part of Gemini programs GN-2008A-Q-25 and GN-2009A-Q-21. 
Some of our 26 target strong lensing clusters have been previously identified as 
strong lenses in the literature: Abell 1703, GHO 132029+315500, RXC J1327.0+0211, 
SDSS J1115+5319, SDSS J1446+3033, SDSS J1527+0652, SDSS J1531+3414, and SDSS J2111-0114 
\citep{hennawi2008}; SDSS J0957+0509, SDSS J1226+2152, SDSS J1621+0607, 
and SDSS J2238+1319 \citep{wen2009}; SDSS J1209+2640 
\citep{ofek2008}; SDSS J1343+4155 \citep{diehl2009,wen2009}; SDSS J1038+4849 
\citep{belokurov2009,kubo2009}; SDSS J2243-0935 \citep{horesh2010}; and 
SDSS J0915+3826 \citep{bayliss2010}. 
Several detailed studies of the strong lensing properties of Abell 1703 can 
be found in the literature \citep{limousin2008,oguri2009b,richard2009,richard2010}.
The remaining $9/26$ of the clusters discussed in this paper are previously 
unpublished strong lenses. 

Analyses of a subset of the Gemini spectroscopy presented 
here have been published in several recent papers. \citet{oguri2009b} conducted
a weak lensing analysis of SDSS J2111-0114, SDSS J1446+3033, SDSS
J1531+3414, and Abell 1703. Gemini spectroscopic redshifts of the
clusters and lensed images were used as constraints in a joint strong
plus weak lensing analysis. Our more careful analysis of these
clusters has revealed additional redshifts of candidate lensed
background sources. \citet{bayliss2010} published the discovery of
two bright, strongly lensed Lyman-$\alpha$ Emitting galaxies at
$z\sim5$ in SDSS J1343+4155 and SDSS J0915+3826 in
\citet{bayliss2010}. In addition, \citet{koester2010} presented the
discovery two bright strongly lensed Lyman Break Galaxies at $z\sim3$
lensed by SDSS J1527+0652 and SDSS J1226+2152.

\begin{deluxetable*}{lllll}
\tablecaption{Summary of Spectroscopic Observations\label{tabobs}}
\tablewidth{0pt}
\tabletypesize{\tiny}
\tablehead{
\colhead{Target} &
\colhead{$\alpha$\tablenotemark{a}} &
\colhead{$\delta$\tablenotemark{a}} &
\colhead{Semester\tablenotemark{b}} &
\colhead{Exposures }
}
\startdata
Abell 1703\tablenotemark{c} & 13:15:05.28 & +51:49:02.9 & 2008A & $2\times2400$s \\
RXC J1327.0+0211\tablenotemark{d,e} & 13:27:01.01 & +02:12:19.5 & 2008A & $2\times2400$s~~~~~~ \\
SDSS J0915+3826 & 09:15:39.01 & +38:26:58.5 & 2008A & $2\times2400$s~~~~~~ \\
SDSS J0957+0509 & 09:57:39.19 & +05:09:31.9 & 2008A & $2\times1200$s, $1\times780$s\tablenotemark{f} \\
SDSS J1115+5319 & 11:15:14.85 & +53:19:54.3 & 2008A & $1\times2400$s\tablenotemark{g}~~~~~~ \\
SDSS J1209+2640\tablenotemark{h} & 12:09:23.69 & +26:40:46.7 & 2008A & $1\times2400$s\tablenotemark{g}~~~~~~ \\
SDSS J1343+4155 & 13:43:32.85 & +41:55:03.5 & 2008A & $2\times2400$s~~~~~~ \\
SDSS J1446+3033 & 14:46:33.45 & +30:33:05.1 & 2008A & $3\times2400$s~~~~~~ \\
SDSS J1527+0652 & 15:27:45.82 & +06:52:33.6 & 2008A & $2\times1200$s~~~~~~ \\
SDSS J1531+3414 & 15:31:10.60 & +34:14:25.0 & 2008A & $3\times2400$s~~~~~~ \\
SDSS J2111-1114 & 21:11:19.34 & -01:14:23.5 & 2008A & $3\times2400$s, $1\times540$s\tablenotemark{f} \\
SDSS J2238+1319 & 22:38:31.31 & +13:19:55.9 & 2008A & $2\times2400$s~~~~~~ \\
-- & -- & -- & -- & -- \\
SDSS J0851+3331 & 08:51:38.87 & +33:31:06.1 & 2009A & $2\times2400$s~~~~~~ \\
SDSS J1028+1324 & 10:28:04.11 & +13:24:52.2 & 2009A & $2\times2400$s, $1\times1560$s\tablenotemark{f} \\
SDSS J1038+4849 & 10:38:43.58 & +48:49:17.7 & 2009A & $2\times2400$s~~~~~~ \\
RCS2 J1055+5548 & 10:55:04.60 & +55:48:23.4 & 2009A & $2\times2400$s~~~~~~ \\
SDSS J1138+2754 & 11:38:08.95 & +27:54:30.7 & 2009A & $2\times2400$s~~~~~~ \\
SDSS J1152+0930 & 11:52:47.39 & +09:30:14.8 & 2009A & $2\times2400$s~~~~~~ \\
SDSS J1152+3313 & 11:52:00.15 & +33:13:42.1 & 2009A & $2\times2400$s~~~~~~ \\
SDSS J1209+2640\tablenotemark{h} & 12:09:23.69 & +26:40:46.7 & 2009A & $2\times2400$s~~~~~~ \\
SDSS J1226+2149\tablenotemark{e,i} & 12:26:51.11 & +21:49:52.3 & 2009A & $2\times2400$s~~~~~~ \\
SDSS J1226+2152\tablenotemark{i} & 12:26:51.69 & +21:52:25.4 & 2009A & $2\times2400$s~~~~~~ \\
GHO 132029+315500\tablenotemark{j} & 13:22:48.77 & +31:39:17.8 & 2009A & $2\times2400$s~~~~~~ \\
SDSS J1420+3955 & 14:20:40.38 & +39:55:10.6 & 2009A & $2\times2400$s~~~~~~ \\
SDSS J1456+5702 & 14:56:00.86 & +57:02:20.6 & 2009A & $2\times2400$s~~~~~~ \\
SDSS J1621+0607 & 16:21:32.37 & +06:07:19.1 & 2009A & $2\times2400$s~~~~~~ \\
SDSS J2243-0935\tablenotemark{d,e} & 22:43:19.80 & -09:35:30.9 & 2009A & $2\times2400$s~~~~~~ \\
\enddata
\tablenotetext{a}{~Coordinates are BCG centroids (J2000.0) calibrated against the SDSS.}
\tablenotetext{b}{~Details of the instrument configuration for each semester can be found in Section 2.3}
\tablenotetext{c}{~This cluster was first identified by \citet{abell1989}.}
\tablenotetext{d}{~Cluster appears in the ROSAT all-sky bright source catalog \citep{voges1999}}
\tablenotetext{e}{~Also a MACS cluster \citep{macs2001}.}
\tablenotetext{f}{~Some N\&S exposure sequences were terminated partway through due to deteriorating conditions at the telescope.}
\tablenotetext{g}{~We have only one N\&S science exposure for SDSS J1209+2640 and SDSS J1115+5319, limiting our ability to correct for chip gaps, chip defects, charge traps, and cosmic rays.}
\tablenotetext{h}{~SDSS J1209+2640 was observed in both semesters with two different masks.}
\tablenotetext{i}{~SDSS J1226+2152 and SDSS J1226+2149 are two strong lensing cores in a larger complex structure. One mask for each core was designed from the same pre-imaging data.}
\tablenotetext{j}{~Cluster first published by \citet{gunn1986}.}
\end{deluxetable*}

\subsection{Imaging}

We obtained pre-imaging of 20 of the 26 clusters in $gri$ with the Gemini Multi-Object 
Spectrograph \citep[GMOS;][]{GMOS} in queue mode in order to facilitate mask design. The 
Gemini imaging data consist of $2\times150$s dithered exposures, with one exposure at
an initial pointing position for a given cluster, and the other exposure at a position 
corresponding to the ``nod'' in our planned spectroscopic observations (see Section 2.3). 
All GMOS images were taken with the detector binned $2\times2$ for a scale of 
$0.1454$ \arcsec~pixel$^{-1}$. Gemini/GMOS-North imaging data was reduced using the
Gemini IRAF\footnote{IRAF (Image Reduction and Analysis Facility) is distributed
by the National Optical Astronomy Observatories, which are operated by AURA,
Inc., under cooperative agreement with the National Science Foundation.} package. 
The pre-imaging have approximate point source $3-\sigma$ limiting magnitudes of 
$g\lesssim 25.5$, $r\lesssim 25.8$, and $i\lesssim 25.5$. 
For four of the 26 clusters -- SDSS J2111-0114, SDSSJ 1446+3033, 
SDSS J1531+3414, and Abell 1703 -- we have only $r-$band pre-imaging from Gemini and 
rely on deep $gri$ imaging from Subaru \citep{oguri2009b} to determine color 
information for sources in these fields. Photometric catalogs for the 24 clusters with 
pre-imaging were derived from the available multi-band imaging data using 
object-finding and aperture photometry routines from the DAOPHOT Package. 
For the remaining two clusters -- SDSS J0957+0509 and SDSS J1527+0562 -- we have 
$g-$band imaging from the 2.5m Nordic Optical Telescope (NOT) on La Palma and the 
3.5m WIYN Telescope on Kitt Peak, respectively. The $g-$band data from NOT consist of 
$2\times300$s exposures taken with the MOSaic CAmera (MOSCA), which is an array of 
$4$ $2k\times2k$ CCDs. Data were taken binned $2\times2$ resulting in $0.217$ 
\arcsec~pixel$^{-1}$.  The $g-$band data from WIYN are similar; we took $2\times300$s 
exposures with the Orthogonal Parallel Transfer Imaging Camera (OPTIC), which is an 
array of $2$ $2k\times4k$ CCDs. These data were unbinned for a scale of $0.14$ 
\arcsec~pixel$^{-1}$. The deeper $g$-band images were used to place slits targeting 
the bright arcs manually, and photometric catalogs from the SDSS DR7 \citep{sdssdr7} 
were used to identify cluster member galaxies by their presence on the red sequence.

\subsection{Mask Design \& Spectroscopy}

Spectroscopic masks for each cluster were designed using object positions and colors 
from the photometric catalogs. The highest priority slits were manually 
placed on candidate lensed background sources as identified by color and morphology, 
and then the mask was filled in with lower priority slits placed on red sequence 
selected cluster members with $r_{AB}\leq22.5$ in the photometric catalogs. This 
flux limit corresponds to a luminosity limit of $\sim0.1 - 0.6$L$^{*}$ for each 
cluster, depending on the cluster redshift.

All spectroscopic observations were carried out with GMOS using the custom 
slitmasks described above. Spectra were taken using the macroscopic 
nod-and-shuffle (N\&S) mode available on GMOS. The use of macroscopic N\&S allows 
for small slits and increases the density of slits that we can place in the cores 
of the  target clusters. We use a modified version of the standard macroscopic 
N\&S mode wherein we shuffle the charge by one third of the detector along the 
spatial axis, while nodding the telescope on the sky by one sixth of the detector. 
A mask is designed to cover the central third of the detector that is effectively 
a combination of two ``sub-masks'', each of which primarily targets a region on the 
sky approximately one sixth the size of the detector. With the nod distance set to 
one sixth of the detector size, the targeted region on the sky is nodded from one 
spectroscopic sub-mask to the other, such that we collect science spectra for this 
region during $100\%$ of the total exposure time of the observation. Our 
strategy avoids the $50\%$ overheads that are necessary for a simple macroscopic 
band N\&S observation by enabling us to design two independent masks covering the 
central sixth of the detector along the spatial axis. The two sub-masks are 
optimized to place slits on as many candidate strong lensing features as possible, 
with slits often placed on the most prominent arcs in both sub-masks to gather 
data on those sources for the full exposure time of the observations. Additionally, 
the sub-masks can include slits targeting sources located in an area equal to 
the size of one sixth of the detector to either side of the central region. These 
regions include red sequence selected cluster members that we use to fill in 
gaps in the slitmask after placing slits on all candidate strong lensing features. 
Figure~\ref{nodandshuffle} shows the Gemini/GMOS $r-$band pre-imaging data for 
SDSS J1138+2754 with the N\&S spectroscopic mask slits over-plotted and each of 
the pointing and nod positions to illustrate our observing strategy.

N\&S offers several benefits that are especially 
advantageous for pursuing redshifts of candidate strongly lensed sources. 
Firstly, N\&S provides for better sky-subtraction 
\citep{glaze2001,abraham2004}, especially at lower spectral resolutions, than traditional 
longslit or multi-slit spectroscopy. Excellent sky-subtraction over a large 
range of wavelengths is crucial for identifying galaxy redshifts at 
$z\gtrsim1.0$, which often relies on spectral lines that are redshifted 
into the red (i.e. $\sim7000-10000$\AA) where sky lines are numerous. 
Secondly, a macroscopic N\&S strategy allows us 
to cut slits matching the sizes of target sources, as small as 
$1\arcsec\times1\arcsec$ microslits, which can be densely packed into the 
cores of our strong lensing clusters to target as many arcs, arclets, and 
cluster members as possible. Our modified N\&S approach 
complements the size of the GMOS detector very nicely. The GMOS detector array 
is approximately $\sim5.6\arcmin \times 5.6$\arcmin in size; this means that we can 
optimize slit placement in the central $\sim1$\arcmin of the target clusters, 
which corresponds well with the core regions probed by strong lensing.

All spectra taken as a part of the GN-2008A-Q-25 program used the R150\_G5306 
grating in first order with the detector binned $2\times2$, producing an average 
dispersion of $3.5$\AA~ per image pixel and a six pixel spectral resolution 
element. The resulting spectral FWHM is $\sim940$ km s$^{-1}$, corresponding to 
a spectral resolution, $R \equiv \frac{\lambda}{\delta \lambda} \simeq 320$, 
and cover a spectral range, $\Delta \lambda \sim4000-9500$\AA, with our highest 
sensitivity in the interval, $\Delta \lambda \sim 5500-9000$\AA. Our effective spectral 
range is limited at both the blue and red ends by the sensitivity of both the GMOS 
CCDs and the transmission efficiency of the grating. The masks for GN-2008A-Q-25 
spectroscopy were designed using only slitlets of $1\arcsec \times 1$\arcsec, many 
of which could be placed along the longest arcs. The N\&S cycle time for all 
2008A spectra was 60s.

Analysis of the GN-2008A-Q-25 spectra motivated us to change the instrumental 
setup for GN-2009A-Q-21 spectroscopy. We no longer
restricted ourselves to only $1\arcsec \times 1 \arcsec$ microslits, but 
instead increased the spatial extent of our slits along the arcs and
occasionally tilted them to better cover an arc or achieve optimal
slit packing. All spectra taken in the GN-2009A-Q-21 
program used the R400\_G5305 grating in first order, in conjunction with the 
GG455\_G0305 longpass filter, and with the detector binned by $2$ in the spectra 
direction and unbinned spatially. This configuration produces a dispersion of 
$1.34$\AA~ per (binned) spectral pixel and spectral resolution of $\sim310$ km s$^{-1}$ 
or $R\simeq960$ with a wavelength coverage, 
$\Delta \lambda \sim4200$\AA. The observed wavelength range for slits located 
near the centers of our masks is $\sim5200-9400$\AA. Given the performance 
of the GMOS CCDs at the very blue and red ends we find that the R400\_G5305 
grating loses very little effective wavelength coverage compared to the 
R150\_G5306, while improving the quality of the N\&S sky subtraction, as well as our 
ability to measure reliable absorption line redshifts for arcs located in the redshift 
desert. 

A persistent problem in our 2008A observations were systematics in the
N\&S sky subtraction caused by charge traps. The amount of trapped
charge depends sensitively on the detector binning and the amount of
charge shuffling (i.e. the N\&S cycle length).  After conducting
experiments with dark frames we found that the detector binned by $2$
in the spectral direction and unbinned spatially provided the best
compromise between trapped charge and increased read noise. We also
experimented with two different N\&S cycle lengths, 60s and
120s, the former of which optimizes sky subtraction and the later of which 
minimizes the negative impact of charge traps.  We obtained test observations 
for one of our masks with both cycles lengths, and found that the quality of the 
N\&S sky subtraction was not significantly diminished for the 120s cycles, which 
we opted to use throughout the remainder of our 2009A observations. 

Table \ref{tabobs} shows which targets were observed from Gemini North
in 2008A and 2009A, along with the integration times for each
mask. Sky subtraction of N\&S data is achieved by simply differencing
the two shuffled sections of the detector. All of our spectra were
wavelength calibrated, extracted, stacked, flux normalized, and
analyzed using a custom data reduction pipeline which we developed
based on the XIDL\footnote{http://www.ucolick.org/$\sim$xavier/IDL/index.html} and
the SDSS idlspec2d\footnote{http://www.astro.princeton.edu/~schlegel/code.html} 
software packages. We extract individual spectra and perform all stacking in 
1D using a rejection algorithm to exclude cosmic rays and hot pixels.  Our 
masks were not observed at the parallactic angle, and we relied on archival
standards in the GMOS data archive to determine the sensitivity
function, thus our flux calibration is only approximate.

In addition to the Gemini/GMOS-North spectroscopy, we supplement our dataset 
with cluster member redshift measurements made at the 3.5m Astrophysical 
Research Consortium (ARC) Telescope at Apache Point Observatory in New Mexico, 
using the Dual Imaging Spectrograph (DIS) in longslit mode. The APO+DIS 
observations were taken using the B400 and R300 gratings on the red and blue 
sides, respectively, and a $1.5\arcsec$ slit. Science exposures were accompanied by 
HeNeAr arc calibrations and quartz lamp flatfield exposures at the same 
orientation in order to minimize systematics errors due to instrument flexure. 
The resulting data were reduced, calibrated, sky-subtracted, extracted and 
stacked using custom IDL scripts that incorporate procedures from the XIDL 
software package. All redshifts measured from the APO+DIS data came out of the red 
side spectra, which cover an wavelength range $\Delta \lambda \simeq 5500-9500$\AA~ 
at a dispersion of $\sim2.3$\AA~pixel$^{-1}$, resulting in spectral resolution 
$R\simeq1100$. We observed RCS2 J1055+5547 on the night of March 17, 2007 
at two different orientations 
selected to simultaneously put $1-2$ bright red sequence selected cluster member 
candidates and $1-2$ arc candidates within the slit. At each orientation we 
collected $3\times900$s integrations and from these data we measure redshifts for the 
brightest cluster galaxy (BCG), which is also present in the SDSS DR7 spectroscopic 
catalog \citep{sdssdr7}, as well as redshifts for two additional cluster members. 
On the night of June 3, 2008 we observed SDSS J1621+0607 at a single orientation 
with $3\times1800$s integrations, from which we measure a redshift for the BCG.

\begin{deluxetable*}{lcllllc}
\tablecaption{Individual Lensed Sources\label{arcproperties}}
\tablewidth{0pt}
\tabletypesize{\tiny}
\tablehead{
\colhead{Cluster Core} &
\colhead{Source Label\tablenotemark{a}} &
\colhead{$redshift$} &
\colhead{$l/w$\tablenotemark{b}} &
\colhead{R$_{arc}$ \tablenotemark{c}} &
\colhead{AB Mag} &
\colhead{Classification\tablenotemark{d}} 
}
\startdata
SDSS J0851+3331 &  A  & $1.6926$ & $16$ & $23\arcsec$ & $g=21.88$ & primary \\
SDSS J0851+3331 &  B  & $1.3454$ & $6$ & ... & $g=24.08$ & secondary \\
SDSS J0851+3331 &  C  & $1.2539$ & ... & ... & $g=23.46$ & tertiary \\
SDSS J0915+3826 &  A  & $1.501$ & $8$ & $12\arcsec$ & $g=22.60$ & primary  \\
SDSS J0915+3826 &  B  & $5.200$ & $4$ & ... & $i=23.34$\tablenotemark{e} & secondary \\
SDSS J0915+3826 &  C  & $1.4358$ & ... & ... & $g=24.85$ & tertiary \\
SDSS J0957+0509 &  A  & $1.8198$ & $11$ & $8\arcsec$ & $g=20.69$ & primary  \\
SDSS J0957+0509 &  B  & $1.0067$ & ... & ... & $g=22.79$ & tertiary  \\
SDSS J0957+0509 &  C  & $1.9259$ & ... & ... & $g=22.29$ & tertiary  \\
SDSS J1038+4849 &  A  & $2.198$ & $14$ & $12\arcsec$ & $g=21.24$ & primary  \\
SDSS J1038+4849 &  B  & $0.9657$ & $12$ & $11\arcsec$ & $g=21.28$ & primary  \\
SDSS J1038+4849 &  C  & $2.783$ & $10$ & $8.5\arcsec$ & $g=22.48$ & primary  \\
SDSS J1038+4849 &  D  & $0.8020$ & $8$ & ... & $r=23.55$ & secondary \\
RCS2 J1055+5547 &  A  & $1.2499$ & $15$ & $16\arcsec$ & $g=22.33$ & primary  \\
RCS2 J1055+5547 &  B  & $0.9359$ & $3$ & ... & $g=23.01$ & secondary \\
RCS2 J1055+5547 &  C  & $0.7769$ & ... & ... & $r=21.31$ & tertiary \\
SDSS J1115+5319 &  D  & $1.234$ & ... & ... & $g=24.63$ & tertiary  \\
SDSS J1138+2754 &  A  & $0.9089$ & $7$ & $9\arcsec$ & $g=21.44$ & primary  \\
SDSS J1138+2754 &  B  & $1.3335$ & $17$ & ... & $g=21.84$ & secondary \\
SDSS J1138+2754 &  C  & $1.455$ & $17$ & ... & $g=23.65$ & secondary \\
SDSS J1152+3313 &  A  & $2.491$ & $13$ & $8.5\arcsec$ & $g=20.84$ & primary  \\
SDSS J1152+3313 &  B  & $4.1422$ & $1$\tablenotemark{f} & ... & $r=23.40$ & secondary  \\
SDSS J1152+0930 &  A  & $0.8933$ & $5$ & ... & $g=22.15$ & secondary \\
SDSS J1152+0930 &  B  & $0.9760$ & ... & ... & $g=24.28$ & tertiary \\
SDSS J1209+2640 &  A  & $1.018$ & $21$ & $11\arcsec$ & $g=21.04$ & primary  \\
SDSS J1209+2640 &  B  & $0.879$ & $6$ & ... & $r=23.45$ & secondary \\
SDSS J1209+2640 &  C  & $3.949$ & $7$ & ... & $r=24.77$ & secondary \\
SDSS J1226+2152 &  A  & $2.9233$ & $12$ & $12\arcsec$ & $g=21.61$ & primary  \\
SDSS J1226+2152 &  B  & $1.3358$ & ... & ... & $r=24.24$ & tertiary  \\
SDSS J1226+2152 &  C  & $0.7278$ & ... & ... & $r=23.70$ & tertiary  \\
SDSS J1226+2152 &  D  & $0.7718$ & ... & ... & $r=22.08$ & tertiary  \\
SDSS J1226+2152 &  E  & $0.7323$ & ... & ... & $r=23.64$ & tertiary  \\
SDSS J1226+2149 &  A  & $1.6045$ & $8$ & $20\arcsec$ & $g=22.44$ & primary  \\
SDSS J1226+2149 &  B  & $0.8012$ & $3$ & ... & $g=22.62$ & secondary  \\
SDSS J1226+2149 &  C  & $0.9134$ & $7$ & ... & $g=23.43$ & secondary  \\
SDSS J1226+2149 &  D  & $1.1353$ & ... & ... & $g=24.48$ & tertiary  \\
Abell 1703 &  A  & $0.889$ & $8$ & $5\arcsec$ & $g=21.93$ & primary  \\
GHO 132029+315500 &  B  & $0.8473$  & $4$ & ... & $r=23.27$ & secondary  \\
GHO 132029+315500 &  C  & $1.1513$  & ... & ... & $g=24.16$ & tertiary  \\
GHO 132029+315500 &  D  & $0.8121$  & ... & ... & $g=24.34$ & tertiary  \\
RXC J1327.0+0211 &  A  & $0.991$ & $8$ & $10\arcsec$ & $g=20.73$ & primary  \\
RXC J1327.0+0211 &  B  & $1.476$ & ... &... & $g=23.77$ & tertiary  \\
RXC J1327.0+0211 &  C  & $1.602$ & ... &... & $g=23.10$ & tertiary  \\
SDSS J1343+4155 &  A  & $2.091$ & $25$ & $13\arcsec$ & $g=20.88$ & primary  \\
SDSS J1343+4155 &  B  & $4.994$ & $2$ & ... & $i=23.78$\tablenotemark{e} & secondary \\
SDSS J1343+4155 &  C  & $1.2936$ & ... & ... & $r=24.60$ & tertiary \\
SDSS J1343+4155 &  D  & $0.9516$ & ... & ... & $g=24.20$ & tertiary \\
SDSS J1420+3955 &  A  & $2.161$ & $7$  & $22\arcsec$ & $g=21.85$ & primary  \\
SDSS J1420+3955 &  B  & $3.066$ & $12$ & $35\arcsec$ & $g=21.87$ & primary \\
SDSS J1446+3033 &  A  & $1.006$  & ... & ... & $g=24.11$ & tertiary  \\
SDSS J1446+3033 &  B  & $0.579$  & ... & ... & $g=22.97$ & tertiary  \\
SDSS J1446+3033 &  C  & $1.441$  & ... & ... & $g=24.47$ & tertiary  \\
SDSS J1456+5702 &  B  & $0.8327$ & $7$ & ... & $r=22.54$ & secondary  \\
SDSS J1456+5702 &  C  & $1.141$ & ... & ... & $g=24.49$ & tertiary \\
SDSS J1527+0652 &  A  & $2.760$ & $10$ & $17\arcsec$ & $g=20.90$\tablenotemark{g} & primary  \\
SDSS J1527+0652 &  B  & $1.283$ & ... & ... & $r=22.70$ & tertiary  \\
SDSS J1531+3414 &  A  & $1.096$ & $9$ & $11\arcsec$ & $g=22.32$ & primary  \\
SDSS J1531+3414 &  B  & $1.300$ & $6$ & $13\arcsec$ & $g=22.15$ & primary  \\
SDSS J1531+3414 &  C  & $1.027$ & ... & ... & $g=22.86$ & tertiary \\
SDSS J1621+0607 &  A  & $4.134$ & $8$ & $16\arcsec$ & $r=22.28$ & secondary  \\
SDSS J1621+0607 &  B  & $1.1778$ & $5$ & ... & $r=21.21$ & primary \\
SDSS J2111-1114 &  A  & $2.858$ & $18$ & $11\arcsec$ & $g=21.18$ & primary  \\
SDSS J2111-1114 &  B  & $1.476$ & ... & ... & $g=22.56$ & tertiary \\
SDSS J2111-1114 &  C  & $1.152$ & ... & ... & $g=23.96$ & tertiary \\
SDSS J2238+1319 &  A  & $0.724$ & $15$ & $10\arcsec$ & $g=21.73$ & primary  \\
SDSS J2238+1319 &  B  & $0.980$ & $3$ & ... & $g=24.22$ & secondary \\
SDSS J2243-0935 &  A  & $2.091$ & $12$ & $10\arcsec$ & $g=21.31$ & primary  \\
SDSS J2243-0935 &  B  & $1.3202$ & $4$ & ... & $g=22.65$ & tertiary \\
SDSS J2243-0935 &  C  & $0.7403$ & $6$ & ... & $r=23.20$ & tertiary \\
\enddata
\tablenotetext{a}{~Source labels matching those in Figures~\ref{color1}-\ref{color7} and Table~\ref{tabredshifts}.}
\tablenotetext{b}{~Length-to-width ratios are all estimated from ground-based imaging with variable seeing. In the case of multiple arcs/images, the largest l/w ratio is given.}
\tablenotetext{c}{~R$_{arc}$ here is the mean distance from a giant arc to the BCG of the lensing cluster.}
\tablenotetext{d}{~Primary, Secondary, or Tertiary background source identification, as discussed in Section 3.1}
\tablenotetext{e}{~From \citet{bayliss2010}.}
\tablenotetext{f}{~This object has $l/w=1$, but we spectroscopically confirm multiple images of the source separated by $\sim13$\arcsec.}
\tablenotetext{g}{~From \citet{koester2010}.}
\end{deluxetable*}

\section{Analysis}

\subsection{Redshift Measurements}

All spectra were examined by eye and compared to a variety of spectral line 
lists spanning a broad rest-frame wavelength range. We assigned redshifts to 
individual spectra by identifying a set of lines at a common redshift, fitting 
a gaussian profile to each line to identify the central wavelength for each line, 
and taking the mean redshift of the entire set of lines. Redshifts for cluster 
member galaxies are derived from at least three lines, the most commonly used of 
which are strong stellar photospheric lines that are characteristic of older 
stellar populations (e.g., CaII H\&K~$\lambda3934,3969$\AA, g-band~$\lambda4306$\AA, 
MgI~$\lambda5169,5174,5185$\AA, and NaI~$\lambda5891,5894,5897$\AA). Redshifts for 
putative strongly lensed sources were measured in the same way as the cluster members, 
though the specific lines used varies significantly among the different lensed source 
spectra. A large majority of our strongly lensed sources are very blue in the available 
photometry, implying that they are actively forming stars. Given our spectral coverage 
we expect to observe one or more prominent emission lines (e.g., [OII]~$\lambda3727$\AA, 
H-$\beta~\lambda4862$\AA, [OIII]~$\lambda4960,5007$\AA~ and H-$\alpha~\lambda6563$\AA) 
for star-forming galaxies at $z\lesssim1.5$, with some slight variation from source 
to source depending on the limit in our red coverage for a given science slit. For 
strongly lensed sources at $z\gtrsim 1.5$ we must rely on rest-frame UV features to 
identify redshifts. In some cases we observe Lyman-$\alpha~\lambda1216$\AA~ in 
emission, accompanied by a break in the continuum, but for many sources we measure 
redshifts from systems of UV metal absorption lines, including but not limited to: 
MgII~$\lambda2796,2803$\AA, FeII~$\lambda2344,2372,2384,2586,2600$\AA, 
CIV~$\lambda1548,1551$\AA, SiII~$\lambda1260,1527$\AA, and SiIV~$\lambda1394,1403$\AA. 
Redshift solutions were also checked against spectral templates, namely the 
\citet{shapley2003} lyman break galaxy (LBG) composite spectrum and the Gemini Deep 
Deep Survey composite late-, intermediate-, and early-type spectra \citep{abraham2004}. 

Redshift errors result primarily from a combination of the uncertainty in our wavelength 
calibrations and the statistical uncertainty in the identification of line centers. The 
measured locations of bright sky lines in wavelength calibrated data taken across 
different nights is stable within the calibration uncertainties discussed above, 
indicating that there are no systematic velocity offsets introduced in comparisons of 
data taken on different dates. Typical total redshifts uncertainties in the case
of high signal-to-noise data -- both cluster member galaxies and background
sources -- are $\pm0.0007$ for spectra taken with the R150
grating/2008A data and $\pm0.0003$ for spectra taken with the R400
grating/2009A data. Lower signal-to-noise data, including approximately half of the
spectra for strongly lensed sources, tend to have slightly larger uncertainties:
as large as $\pm0.001$ for R150/2008A spectra and $\pm0.0006$ for R400/2009A
spectra. We also note that redshifts for some of our background sources that are measured from 
only a few features in the rest-frame UV can often be subject to additional systematic 
uncertainty due to the inherent velocity offsets that are typically observed between
absorption and emission features in star forming galaxies at high redshift
\citep[e.g.,][]{shapley2003}.

Each redshift measurement falls into one of four classifications ($0 - 3$) which describe 
the confidence level of the redshift. Class $3$ redshifts are the highest confidence 
measurements and are typically measured from systems of $\geq 6$ absorption and emission 
features. These redshift measurements are secure with essentially no chance of 
misinterpretation, and the large majority of the redshifts reported here are of 
this classification. Figure~\ref{gmosspec3} shows examples of six class $3$ spectra. 
Class $2$ redshifts are medium-confidence measurements that are based on at least 
two high-significance lines and/or a larger number of low-significance features. The 
redshifts reported here as class 2 are very likely the real redshifts of the corresponding 
sources, but there is a small chance that any given class $2$ redshift might have been 
mis-identified. Two example class 2 spectra are shown in Figure~\ref{gmosspec2}. Class $1$ 
redshifts are low-confidence measurements that were made using only a few low-significance 
spectral features, and represent a ``best-guess'' redshift using the available spectral 
data along with color information in the pre-imaging data. Figure~\ref{gmosspec1} 
shows two example class 1 spectra.

\begin{figure*}[t]
\centering
\includegraphics[scale=0.9]{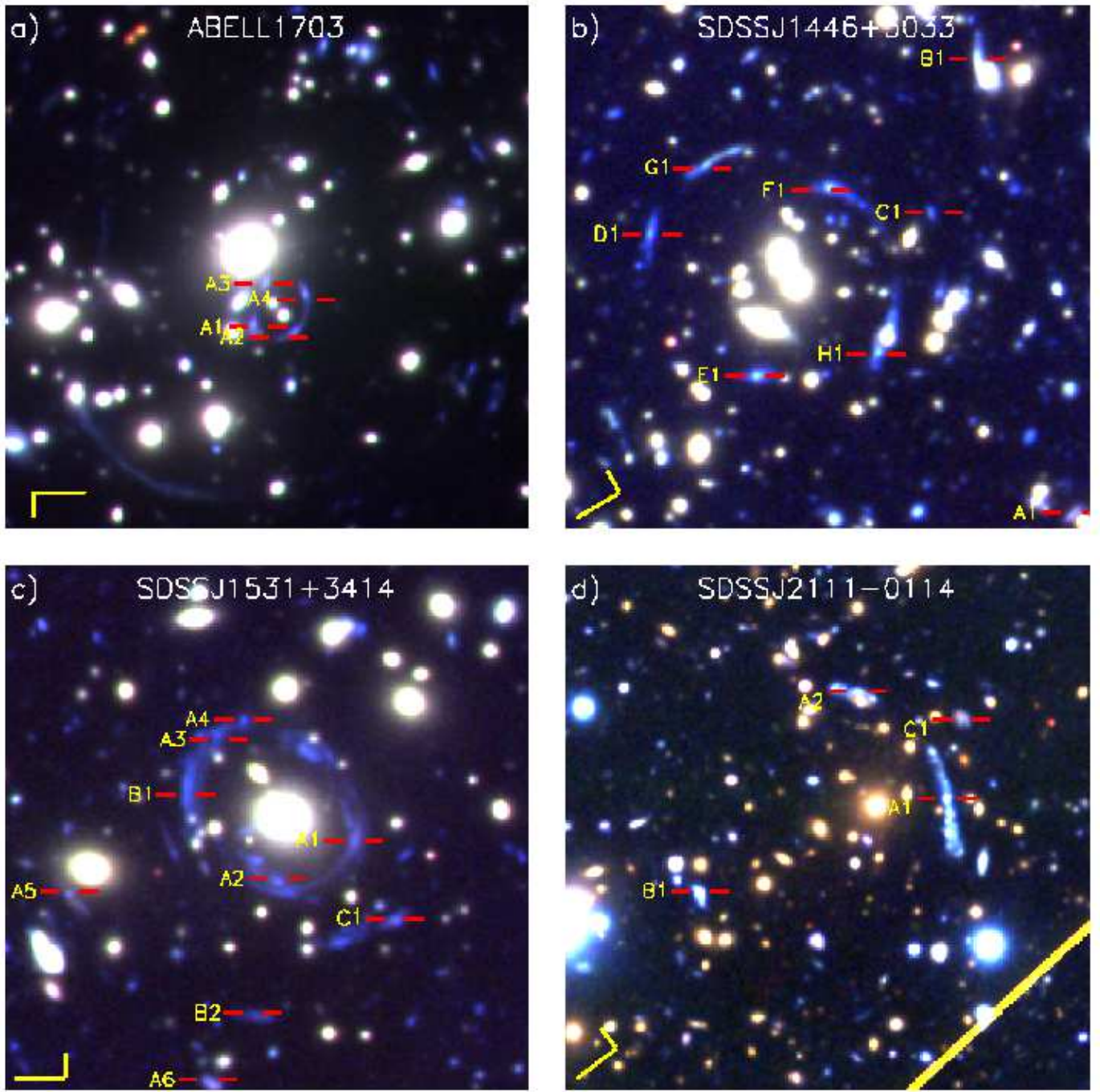}
\caption{\scriptsize{
Target strong lensing cluster fields -- $a)$ Abell 1703 , $b)$ SDSS J1446+3033, 
$c)$ SDSS J1531+3414, and $d)$ SDSS J2111-0114. Color composite 
images are made from $gri$ imaging obtained with Subaru/SuprimeCam 
\citep[see][]{oguri2009b}.  All images are $75\arcsec \times 75\arcsec$. 
Background sources are bracketed by red lines and labeled. Source labels 
with the same letter but different numbers (e.g. A1, A2, etc.) have 
the same redshifts to within the measurement errors, and 
are presumed to be the same source, multiply imaged. Labels 
can be used to match sources in the images with their measured 
redshifts in Table~\ref{tabredshifts}. North and East are indicated by 
the yellow axes in the lower left corner of each image, with North 
being the longer axis.
}}
\label{color1}
\end{figure*}

\begin{figure*}[t]
\centering
\includegraphics[scale=0.9]{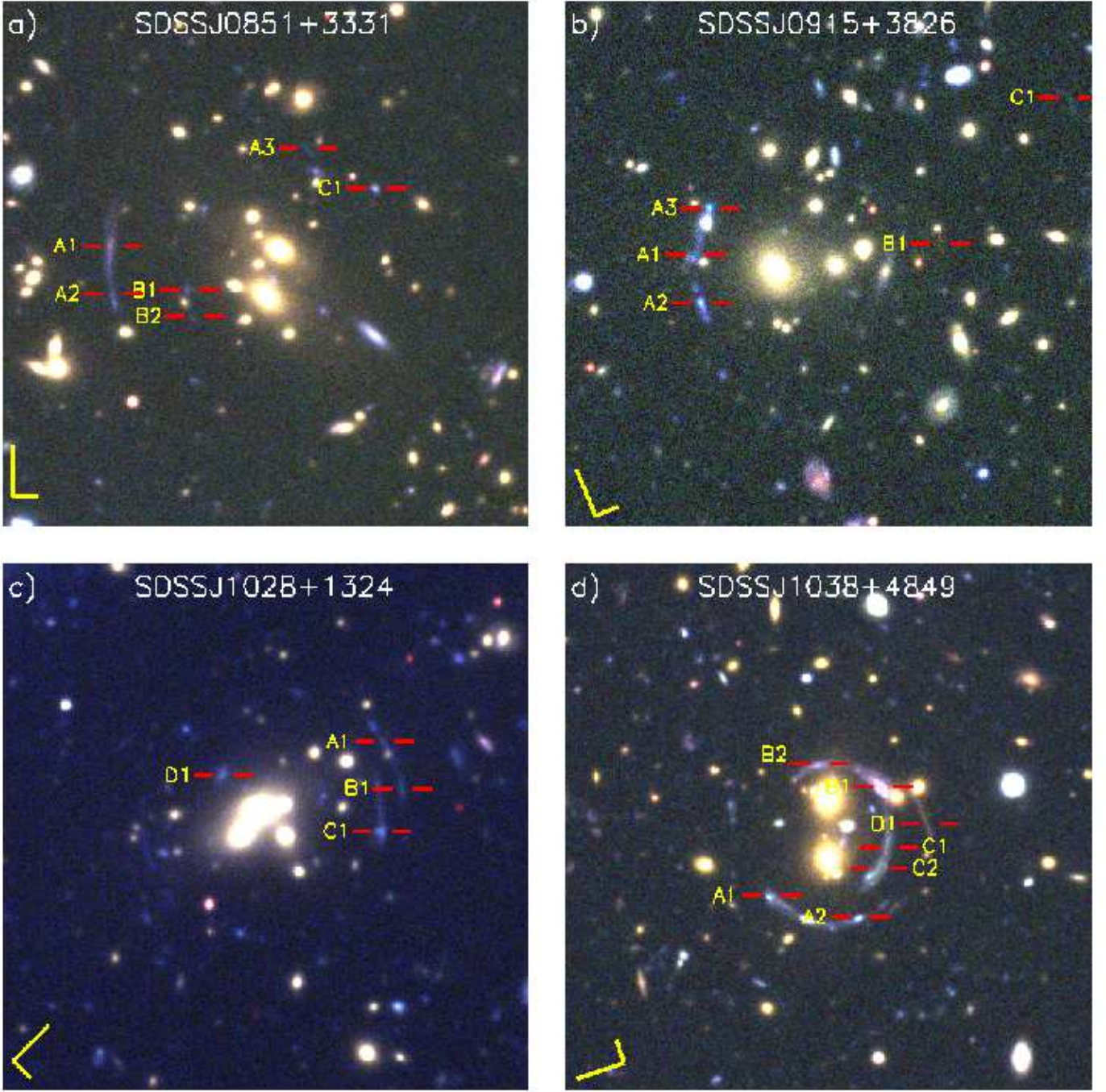}
\caption{\scriptsize{
$a)$ SDSS J0851+3331, $b)$ SDSS J0915+3826, $c)$ SDSS J1028+1324, and 
$d)$ SDSS J1038+4849. Color composite images are made from $gri$ 
pre-imaging data from Gemini/GMOS-North, $75\arcsec \times 75\arcsec$. 
Sources are bracketed and labeled in the same fashion as in Figure~\ref{color1}. 
}}
\label{color2}
\end{figure*}

\begin{figure*}[t]
\centering
\includegraphics[scale=0.9]{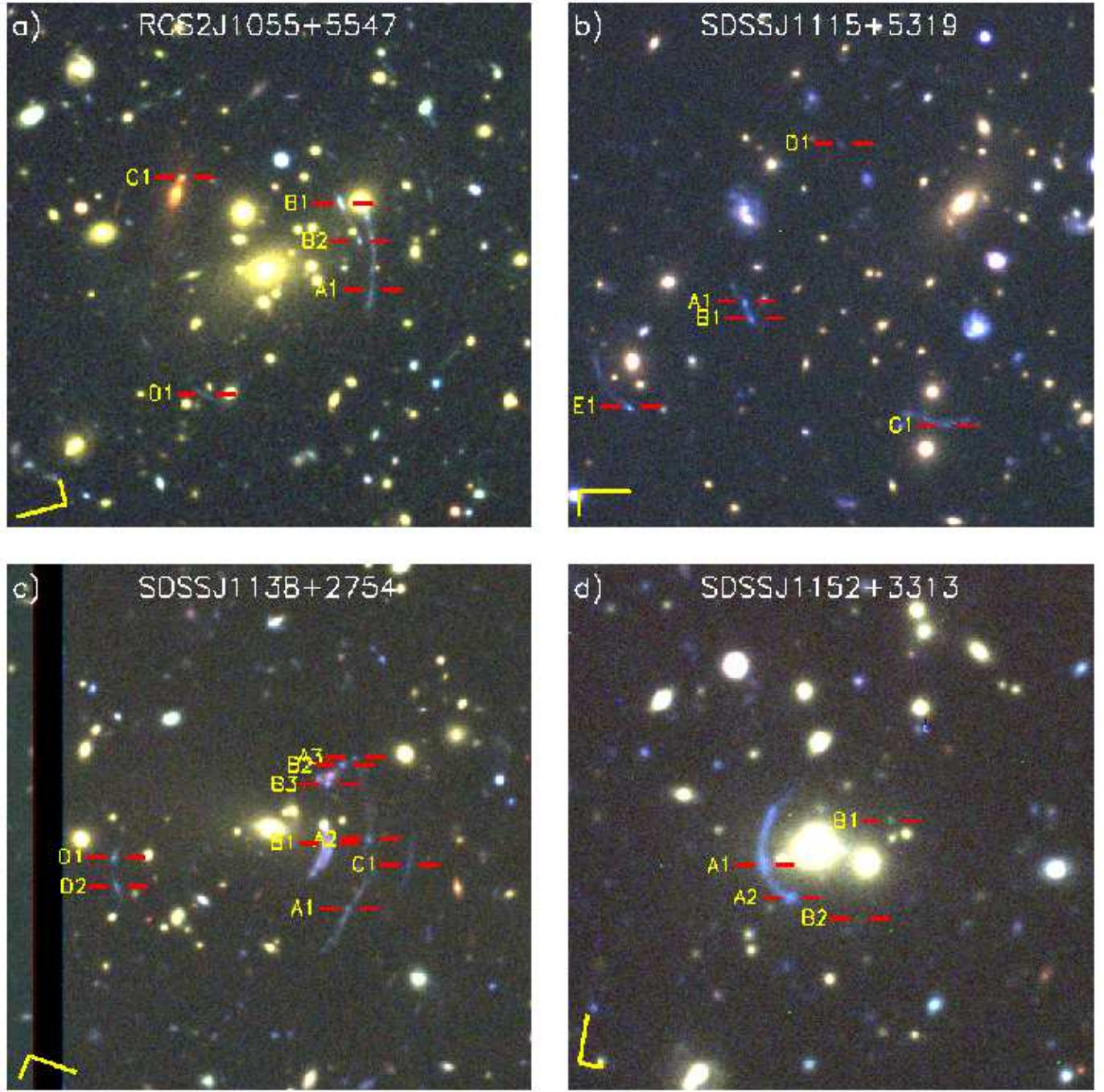}
\caption{\scriptsize{
$a)$ RCS2 J1055+5547, $b)$ SDSS J1115+5319, $c)$ SDSS J1138+2754, and 
$d)$ SDSS J1152+3313. Color composite images are made from $gri$ 
pre-imaging data from Gemini/GMOS-North, $75\arcsec \times 75\arcsec$. 
Sources are bracketed and labeled in the same fashion as in Figure~\ref{color1}.
}}
\label{color3}
\end{figure*}

\begin{figure*}[t]
\centering
\includegraphics[scale=0.9]{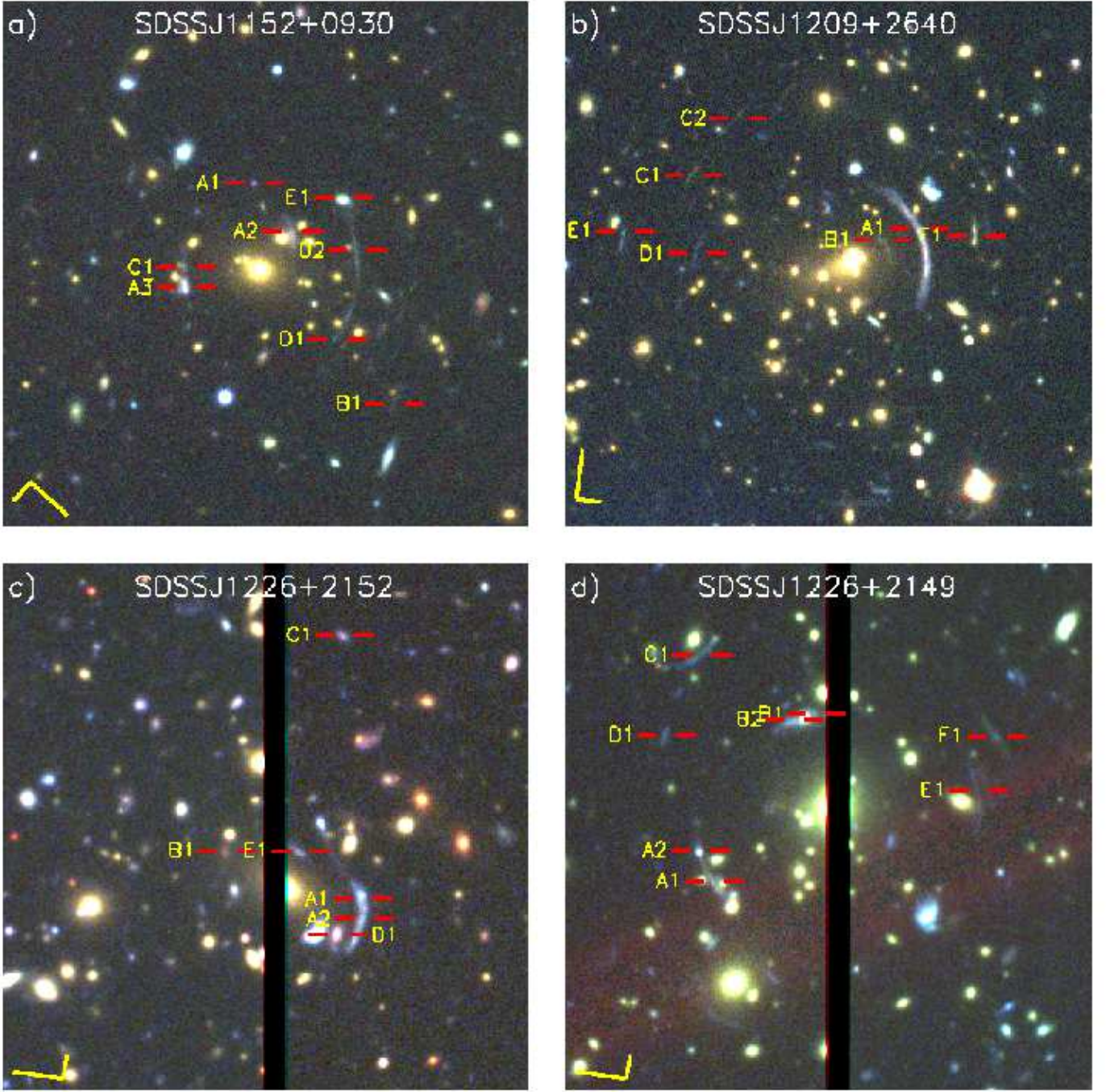}
\caption{\scriptsize{
$a)$ SDSS J1152+0930, $b)$ SDSS J1209+2640, $c)$ SDSS J1226+2152, and 
$d)$ SDSS J1226+2149. Color composite images are made from $gri$ pre-imaging 
data from Gemini/GMOS-North, $75\arcsec \times 75\arcsec$. Sources are 
bracketed and labeled in the same fashion as in Figure~\ref{color1}.
}}
\label{color4}
\end{figure*}

\begin{figure*}[t]
\centering
\includegraphics[scale=0.9]{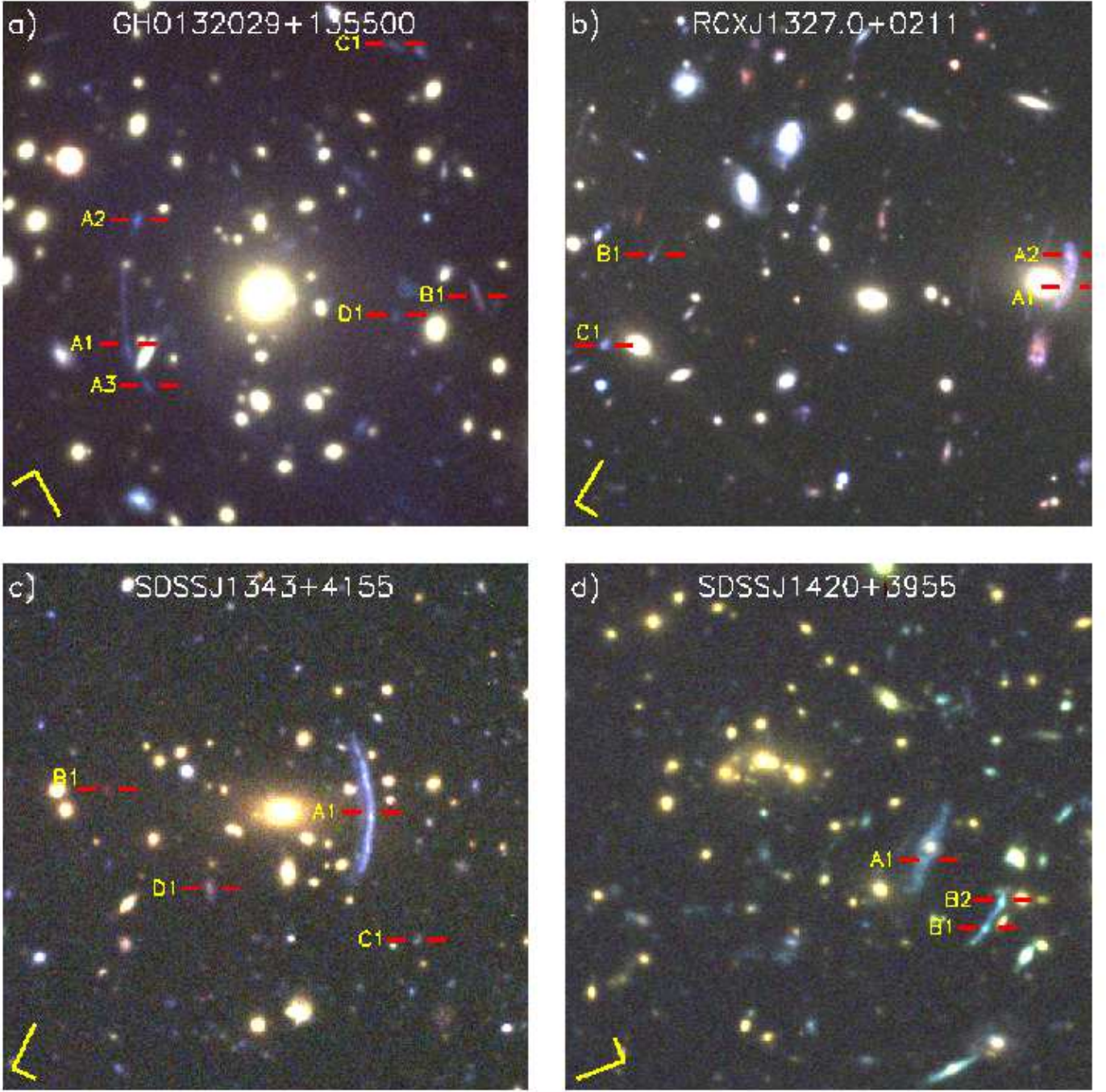}
\caption{\scriptsize{
$a)$ GHO 132029+315500, $b)$ RXC J1327.0+0211, $c)$ SDSS J1343+4155, and 
$d)$ SDSS J1420+3955. Color composite images are made from $gri$ pre-imaging 
 data from Gemini/GMOS-North, $75\arcsec \times 75\arcsec$. Sources are 
bracketed and labeled in the same fashion as in Figure~\ref{color1}. There 
is a triangular region of apparent emission in the color image for 
GHO 132029+315500 (panel a), which is the result of ghosting from a 
bright star located near the cluster on the sky.
}}
\label{color5}
\end{figure*}

\begin{figure*}[t]
\centering
\includegraphics[scale=0.9]{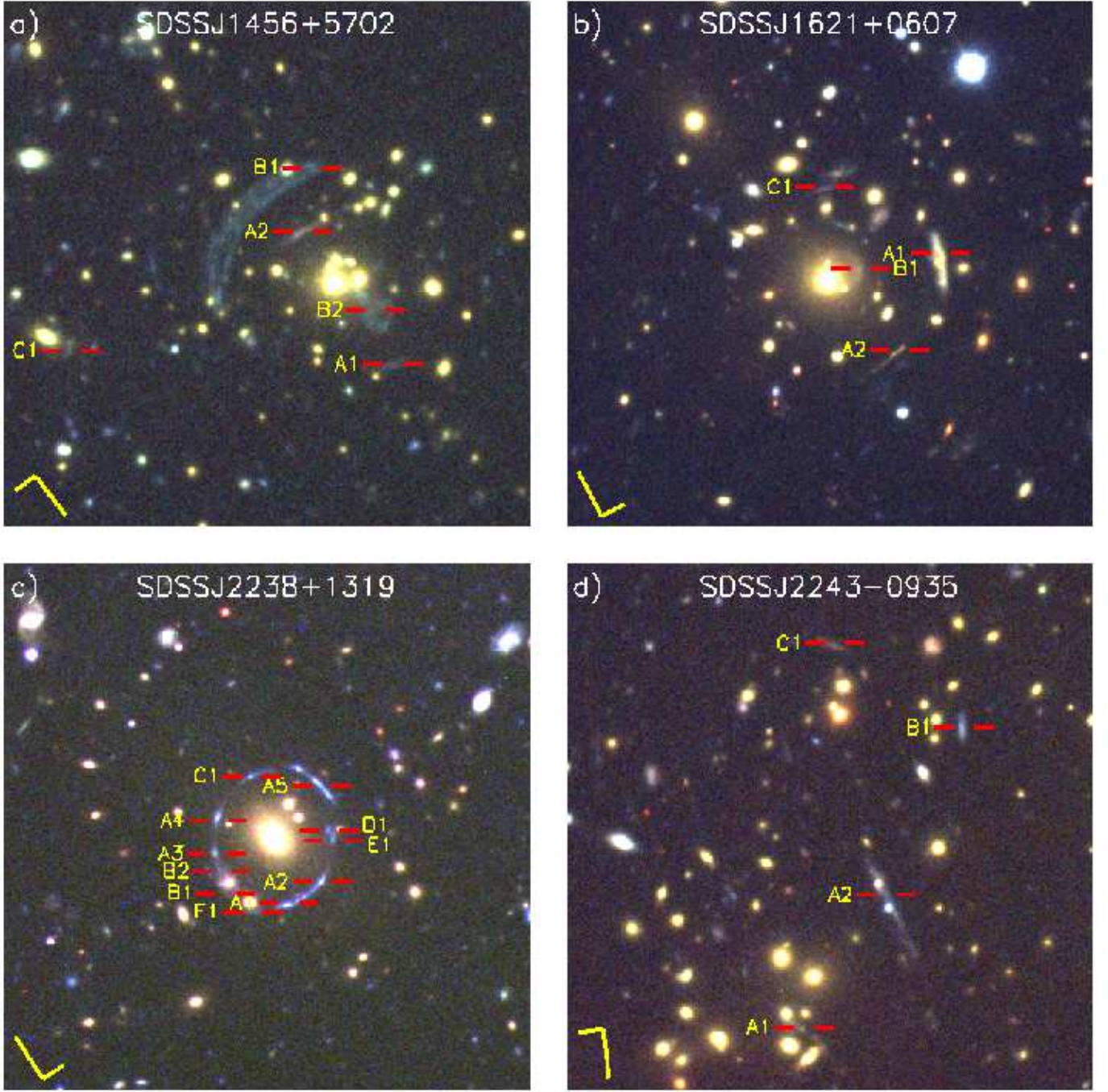}
\caption{\scriptsize{
$a)$ SDSS J1456+5702, $b)$ SDSS J1621+0607, $c)$ SDSS J2238+1319, and 
$d)$ SDSS J2243-0935. Color composite images are made from $gri$ pre-imaging 
data from Gemini/GMOS-North, $75\arcsec \times 75\arcsec$. Sources are 
bracketed and labeled in the same fashion as in Figure~\ref{color1}.
}}
\label{color6}
\end{figure*}

\begin{figure*}[t]
\centering
\includegraphics[scale=0.9]{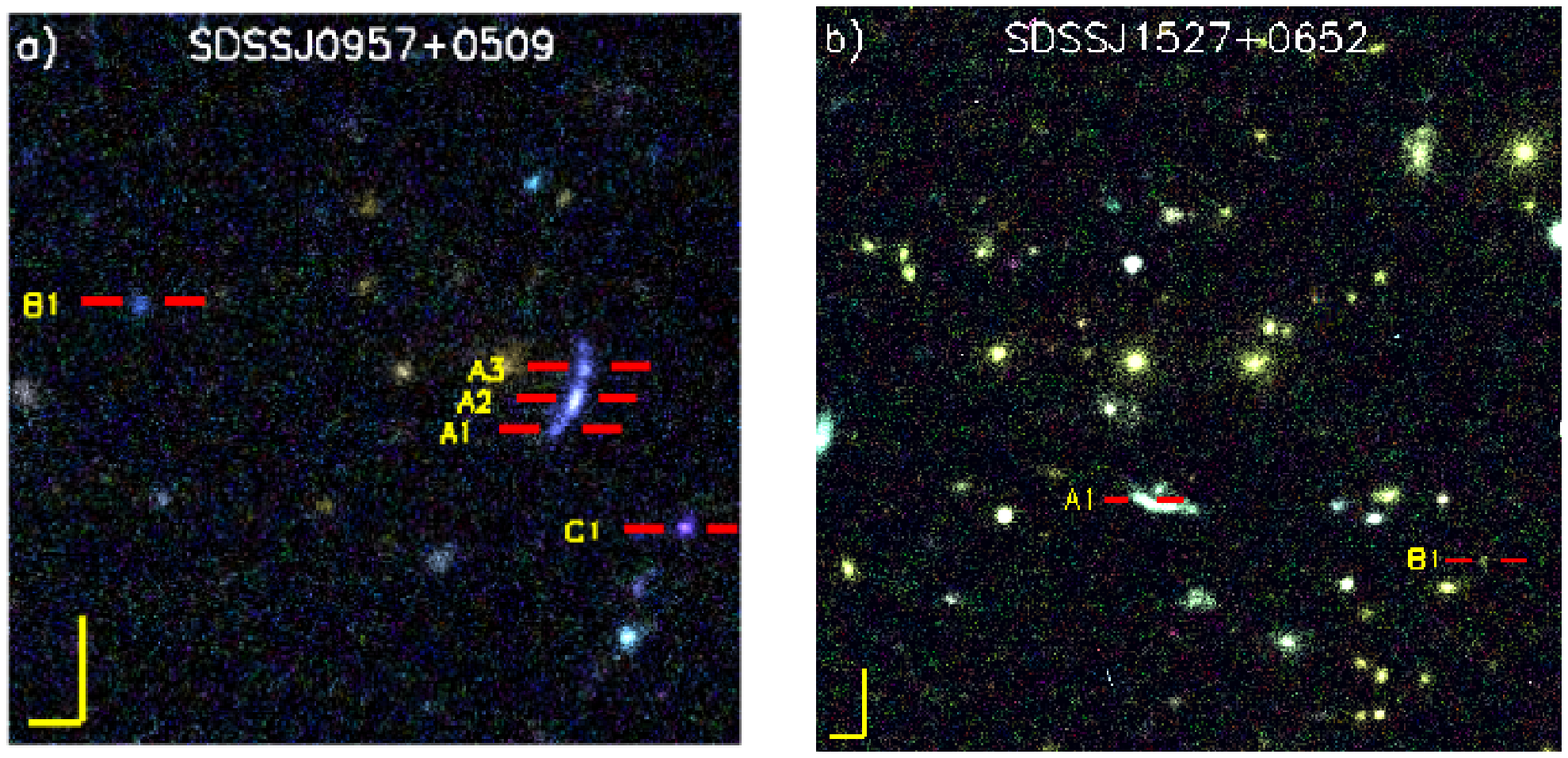}
\caption{\scriptsize{
$a)$ SDSS J0957+0509 and $b)$ SDSS J1527+0652 $75\arcsec \times 75\arcsec$
color composite images are made from $g-$band imaging from the Nordic Optical
Telescope (SDSS J0957+0509) and WIYN Telescope (SDSS J1527+0652), combined
with color information from the SDSS. Multi-object spectroscopy slitmasks
for these two clusters were designed without pre-imaging from Gemini/GMOS.
}}
\label{color7}
\end{figure*}

Class $0$ indicates a redshift failure for a particular slit; some objects labeled 
Class $0$ exhibit low S/N continuum flux but lack sufficiently strong lines or 
dominant features to facilitate a redshift measurement. Class $0$ spectra 
correspond to objects that are good candidates to be strongly lensed background 
sources based on their color, location, and morphology, that were targeted by our 
spectroscopic masks. We report all background source redshifts measured for each of our
$26$ strong lensing clusters, with each redshift tied to a source on the sky by its
foreground cluster name and a two character object label, where the first character
of the label indicates a unique background source and the second character of the
label indicates a slit placed on that background source. All cluster member galaxies 
and background sources with redshifts, as well as Class $0$ candidate strongly lensed 
sources are presented in Table~\ref{tabredshifts} 
with labels that correspond to the label markers in Figure~\ref{color1},
Figure~\ref{color2}, Figure~\ref{color3}, Figure~\ref{color4}, Figure~\ref{color5},
Figure~\ref{color6}, and Figure~\ref{color7}. Figures~\ref{color1}-\ref{color7}  are 
color images of each lensing cluster field with the object labels over-plotted to 
indicate the source locations. In many cases our spectroscopic masks had more 
slits than are indicated in the color images, but we combined spectra for slits 
that were directly adjacent to one another and those slits which contained spectra 
from different pieces of what is clearly the same extended source.

In total our Gemini spectroscopy includes a total of $1126$ science spectra on
$26$ different masks ($\simeq43$ slits per mask). In many cases there are
multiple slits on a mask that target a single background lensed source.
This occurs in some masks where we place slits on sources at both the pointing and
nod positions to collect science spectra for $100\%$ of our exposure time. Most
masks have multiple slits placed on separate images of the same multiply imaged
source, or multiple slits placed along different pieces of a continuous giant arc;
this last case is demonstrated in the mask displayed in
Figure~\ref{nodandshuffle}. In addition to the Gemini/GMOS-North spectroscopy, 
we also present analysis of a few cluster member spectra obtained on the ARC 3.5m 
telescope, with DIS. Redshifts from APO/DIS spectra were measured in the same way 
as the Gemini/GMOS redshifts. Combining all cluster member spectra results in a 
total of $262$ spectroscopic cluster member redshifts. We supplement our own 
measurements with $26$ cluster member redshifts from the SDSS DR7 spectroscopic 
catalog in order to characterize the dynamical properties of the strong lensing 
clusters with an average sample size of $11$ spectroscopic members per cluster.

From the slits placed on candidate strong lensing features we identify $126$ spectra with
redshifts that place them behind the foreground galaxy clusters, and we associate these spectra
with $69$ unique background galaxies, many of which are obviously strongly lensed and/or
multiply imaged, and all of which are likely magnified significantly. We divide these
$69$ individual lensed background sources into three distinct samples: primary giant
arcs, secondary strongly lensed sources, and tertiary background sources. Primary
giant arcs are those giant arcs that were initially used to identify a given cluster
as a strong lens in the SDSS imaging data. There is typically one primary giant arc
per cluster lens, though some systems, such as SDSS J1038+4849 and SDSS J1446+3033, have
multiple, distinct primary giant arcs that are visible in the SDSS survey data. Secondary
strongly lensed sources are objects which either form arcs, or are multiply imaged, such
that we identify them follow-up imaging but lack sufficient brightness and/or morphology 
to be identified as arcs in the raw SDSS survey imaging. Primary and secondary sources are 
likely magnified by factors of $\gtrsim10\times$ 
\citep[e.g.,][]{richard2009,bayliss2010,koester2010}. 
Tertiary background sources are sources or arclets that are located behind one of our 
cluster lens targets but which do not appear to be strongly lensed based on the available 
data. Tertiary background sources are likely magnified by anywhere between a few tens of 
percent and factors of a few due to their location near the core of the foreground cluster 
lenses \citep{smail2002}.

Table~\ref{arcproperties} contains a list of all unique background sources with secure 
redshifts from GMOS spectroscopy. Sources are listed as either primary, secondary
or tertiary objects. Primary giant arcs are listed with measurements of the length-to-width 
(l/w) ratio, the average radial separation between the arc and the cluster center (R$_{arc}$), 
and total integrated AB magnitudes in the $g-$ band, or in one of the $r-$ or $i-$bands if 
a given arc has poor signal-to-noise in our $g-band$ imaging data. We also report 
l/w ratio estimates and integrated AB magnitudes for secondary strongly lensed sources, 
and integrated AB magnitudes for tertiary sources. This table does not include any sources 
for which we do not have precise redshift measurements, and so the primary arcs around 
some clusters -- SDSS J1028+1324, SDSS J1115+5319, SDSS J1152+0930, GHO 132029+315500, 
SDSS J1446+3414, and SDSS J1456+5702 -- do not appear in Table~\ref{arcproperties}. 
Similarly there are dozens of putative secondary strongly lensed sources apparent in 
the GMOS pre-imaging that are not listed in Table~\ref{arcproperties} because the 
spectroscopy did not yield redshifts. Some arcs without precise redshifts are 
addressed in \citet{bayliss2011} and have ``redshift desert'' constraints placed 
on the strongly lensed sources.

The magnitudes given in Table~\ref{arcproperties} are simple integrated aperture 
magnitudes of the brightest contiguous image or arc for a given background source, 
where apertures are drawn by eye to match the morphology of the arcs/sources. The 
photometry is calibrated relative to stars in the SDSS, and are intended only to 
give a rough sense of the brightness for a given source. These magnitude 
measurements have typical errors of $\sim \pm0.1$ magnitudes, and we emphasize 
that the aperture magnitudes can be misleading in some cases. For example, the 
large arc around SDSS J1456+5702 that covers an approximate area on the 
sky of $\sim60-70\arcsec^{2}$ (see Figure~\ref{color6}).

%number 4a

%number 4b

%number 4c

%number 4d

%number 4e

%number 4f

%number 4g

\subsection{Cluster Velocity Dispersions and Dynamical Masses}

\begin{figure*}[t]
\centering
\includegraphics[scale=0.58]{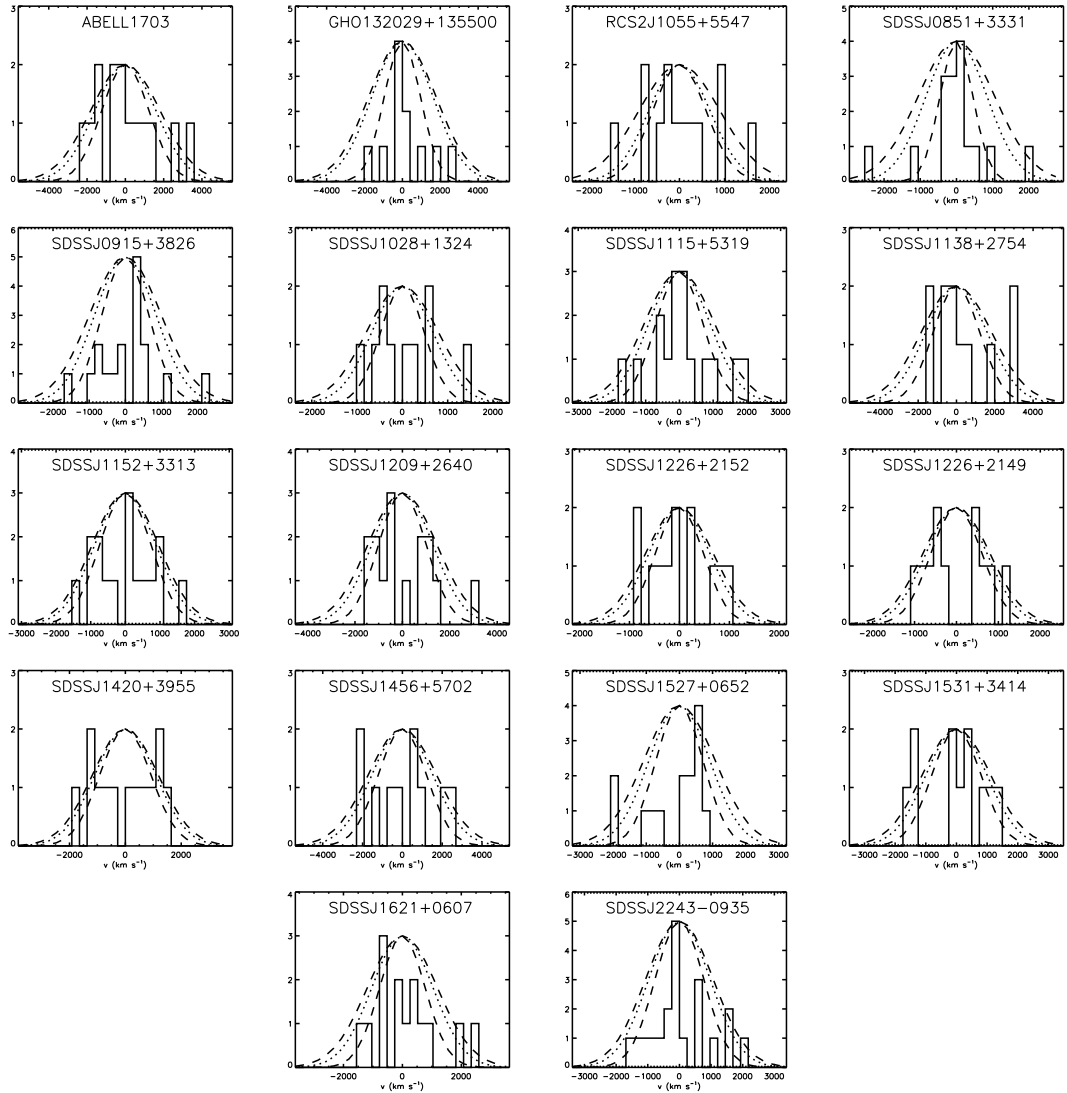}
\caption{\scriptsize{
Velocity histograms for the $18$ clusters with $N\geq10$ cluster 
member redshifts are plotted as histograms. Best-fit gaussians 
with the mean and variance values from the bi-weight estimator 
for each cluster are over-plotted (dotted lines), along with 
the $1-\sigma$ errors on the velocity dispersion (dashed lines). 
}}
\label{sigmavs}
\end{figure*}

Results from spectroscopy of the $26$ cluster lenses are summarized 
in Table~\ref{tabclusters}. There are $18$ clusters in our sample 
with $N_{spec} \geq 10$ spectroscopically confirmed cluster members which 
we take as the minimum number of cluster members that can produce a 
velocity dispersion estimate that is robust against large biases due to 
small sampling. The velocity dispersion of individual galaxies within 
galaxy clusters is a cluster mass observable that has a long history in 
astronomy \citep[e.g.,][]{smith1936,zwicky1937} and remains a viable method 
for estimating the total masses of cluster by its dynamics. Estimates of 
the variance of poorly sampled distributions can be easily biased and require 
algorithms beyond the simple median and standard deviation. We use the bi-weight 
estimator of \citet{beers1990} to determine the redshifts and 
velocity dispersions for our cluster sample, and compute the errors 
on the velocity dispersion by calculating the bi-weight estimate of the 
dispersion for many bootstrapped realizations of the velocity data for 
each cluster and identifying the upper and lower $68\%$ confidence intervals.
Velocity histograms for the $18$ strong lensing clusters with $N\geq10$ 
spectroscopic members are plotted in Figure~\ref{sigmavs}, along with 
best-fit gaussian models. 

Computing the dynamical mass from cluster member velocities requires some 
understanding of the relationship between the velocity dispersion of 
dark matter in the clusters and the velocity dispersion individual 
member galaxies, often parameterized as the velocity bias, 
$b_{v} = \sigma_{gal} / \sigma_{dm}$. Here $\sigma_{gal}$ and 
$\sigma_{dm}$ are the 1-dimensional velocity dispersions of member 
galaxies and dark matter particles, respectively.  Measuring the velocity bias 
is difficult because it requires two independent mass estimates for 
a sample of clusters, one dynamical, and in reality all available mass 
observables are subject to significant systematics and errors. Studies of 
numerically simulated halos can also be used to predict what the 
velocity bias should be for a given population of halos in a given 
cosmology by identifying and tracking the velocities of ``subhalos'' 
within clusters, where the subhalos presumably host cluster member 
galaxies.

\begin{figure}
\centering
\includegraphics[scale=0.38]{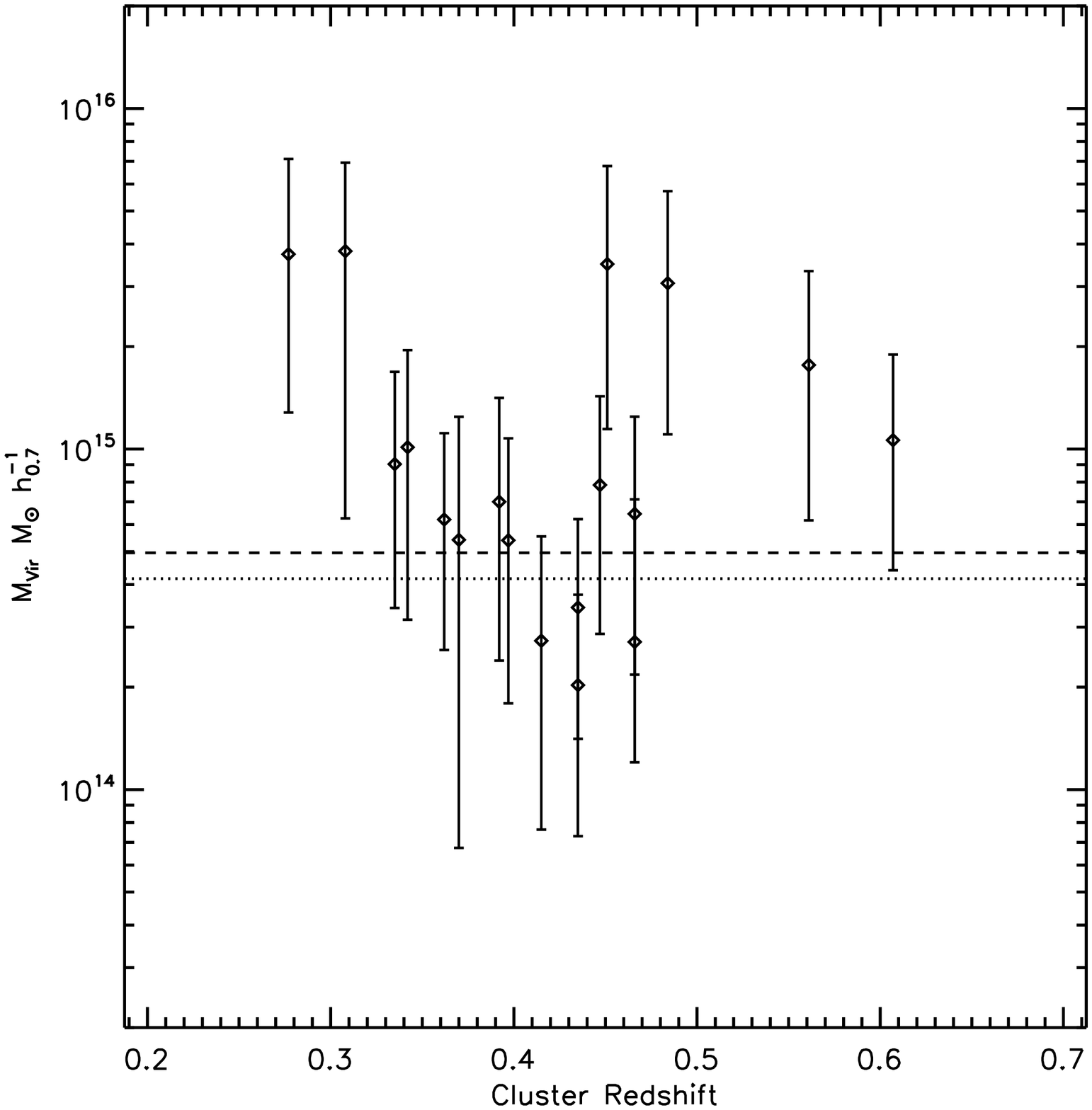}
\caption{\scriptsize{
Dynamical $M_{Vir}$ plotted against the lensing cluster redshift
for each of our observed strong lensing clusters that have $N\geq10$
spectroscopically measured cluster members. Because
arc/arclet candidates were prioritized in our GMOS spectroscopy we
typically have only $\sim14$ confirmed members per cluster, which
limits our ability to estimate $M_{Vir}$ for individual clusters to 
better than an order of magnitude. 
Even so, these rough dynamical mass estimates are sufficient to confirm 
that our strong lensing selected clusters are primarily drawn from
the extreme high-mass end of the halo mass function, and have median 
$M_{Vir} = 7.84 \times10^{14} M_{\sun} h_{0.7}^{-1}$. Over-plotted are 
predictions for the median $M_{Vir}$ of strong lensing selected clusters 
from \citet{hennawi2007} (dotted line), as well as this the predicted 
median $M_{Vir}$ after accounting for the expected $19\%$ bias in 
dynamical masses calculated for strong lensing selected clusters 
(dashed line).}}
\label{dynamicalmass}
\end{figure}

Efforts to make such predictions have produced estimates of 
the velocity bias in the range $b_{v} \sim 1.0-1.3$ 
\citep{colin2000,ghigna2000,diemand2004,faltenbacher2005}. More 
recent work indicates that the way in which subhalos are tracked and 
defined in a simulation effects the resulting velocity bias prediction, 
and studies in which subhalos are treated correctly produce a velocity 
bias that is consistent with little or no significant bias \citep{faltenbacher2006,white2010}. 
Based on these recent results, we assume no velocity bias ($b_{v} = 1$) 
between the galaxy and dark matter velocity dispersion for each cluster in 
our sample. \citet{white2010} also investigated the relationship between 
$\sigma_{gal}$ and $\sigma_{dm}$ for individual simulated halos as a 
function of the number of available 
spectroscopic cluster members. Their results suggest that for the best cases, 
$N_{members}\geq50$, there is an intrinsic scatter of $\sim15\%$ between 
$\sigma_{gal}$ and $\sigma_{DM}$ for a given halo, and that this scatter 
is much worse -- as high as $\sim20\%$ -- when as few as 10 cluster members 
are used. We conservatively fold an additional $20\%$ fractional uncertainty 
into our dynamical mass calculations to reflect the scatter between the 
galaxy velocity dispersion and the true dark matter velocity dispersion 
for our clusters. To calculate $M_{200}$ we apply the $\sigma_{DM} - M_{200}$ 
relation from \citet{evrard2008} for the $18$ of our strong lensing clusters 
with $N\geq10$ spectroscopic members and plot the resulting masses against the 
corresponding cluster redshift in Figure~\ref{dynamicalmass}. Our dynamical data are 
based on small numbers of spectroscopic members and the resulting $M_{200}$ values 
lack precision. However, we can use our data to get a general sense of the mass 
scale of the halos that we are probing with strong lensing selected clusters. 
The expectation is that mass is the dominant property in determining 
the likelihood of a given cluster to produce giant arcs 
\citep[e.g.,][]{hennawi2007}. Observational results comparing the fraction of 
X-ray vs. optically selected clusters which produce giant arcs are consistent 
with this general expectation \citep{horesh2010}.

\begin{deluxetable*}{llcl}
\tablecaption{Properties of the Galaxy Cluster Lenses\label{tabclusters}}
\tablewidth{0pt}
\tabletypesize{\tiny}
\tablehead{
\colhead{Cluster Core Name} &
\colhead{$z$} &
\colhead{$N_{members}$\tablenotemark{a}} &
\colhead{$\sigma_{v}$ (km s$^{-1}$)} }
\startdata
SDSS J0851+3331  &  $0.370$  & $16$  &  $844^{+214}_{-390}$   \\
SDSS J0915+3826  &  $0.397$  &  $17$  &  $846^{+142}_{-200}$   \\
SDSS J0957+0509  &  $0.448$  &  $8$  &  $1006^{+210}_{-270}$   \\
SDSS J1028+1324  &  $0.415$  &  $10$   &  $675^{+120}_{-193}$   \\
SDSS J1038+4849  &  $0.430$  &  $7$   &  $638^{+123}_{-37}$    \\
RCS2 J1055+5547  &  $0.466$  &  $13$  &  $678^{+221}_{-89}$   \\
SDSS J1115+5319  &  $0.466$  &  $16$  &  $907^{+132}_{-210}$   \\
SDSS J1138+2754  &  $0.451$  &  $11$  &  $1597^{+238}_{-384}$   \\
SDSS J1152+3313  &  $0.362$  &  $16$  &  $883^{+74}_{-142}$   \\
SDSS J1152+0930  &  $0.517$  &  $6$   &  $1360^{+110}_{-322}$   \\
SDSS J1209+2640  &  $0.561$  &  $15$  &  $1290^{+166}_{-284}$    \\
SDSS J1226+2149  &  $0.435$  &  $12$  &  $612^{+67}_{-129}$   \\
SDSS J1226+2152  &  $0.435$  &  $11$  &  $730^{+71}_{-119}$   \\
Abell 1703       &  $0.277$  &  $14$\tablenotemark{b}  &  $1597^{+217}_{-362}$   \\
GHO 132029+315500 &  $0.308$  &  $11$  &  $1614^{+158}_{-660}$   \\
RXC J1327.0+0211  & $0.259$  &  $9$   &  $683^{+127}_{-305}$   \\ 
SDSS J1343+4155  &  $0.418$  &  $7$   &  $1011^{+199}_{-287}$   \\
SDSS J1420+3955  &  $0.607$  &  $13$  &  $1095^{+86}_{-175}$   \\
SDSS J1446+3033  &  $0.464$  &  $4$   &  $973^{+149}_{-233}$   \\
SDSS J1456+5702  &  $0.484$  &  $10$  &  $1536^{+183}_{-324}$   \\
SDSS J1527+0652  &  $0.392$  &  $14$  &  $923^{+162}_{-210}$   \\
SDSS J1531+3414  &  $0.335$  &  $11$  &  $998^{+120}_{-194}$   \\
SDSS J1621+0607  &  $0.342$  &  $14$  &  $1038^{+150}_{-265}$   \\
SDSS J2111-0114  &  $0.638$  &  $6$   &  $1192^{+174}_{-339}$  \\
SDSS J2238+1319  &  $0.411$  &  $7$   &  $318^{+26}_{-86}$   \\
SDSS J2243-0930  &  $0.447$  &  $20$  &  $966^{+96}_{-199}$   \\
\enddata
%\label{table_clusters}
\tablenotetext{a}{~Number of spectroscopic cluster members, including galaxies with
spectroscopy pulbically available from in the SDSS DR7.}
\tablenotetext{b}{~Includes 10 additional cluster member redshifts taken 
from various published studies of Abell 1703 \citep{allen1992,rizza2003,richard2009}.}
\end{deluxetable*}

%number 5?

%number 6?

\section{Discussion}

Before we discuss the implications of our data we must note that the sample of
strong lensing clusters that we targeted with Gemini are not drawn randomly from
our full catalog of visually selected cluster lenses in the SDSS. Rather, we have
generally obtained follow-up spectroscopy for strong lensing clusters with
the largest apparent giant arc radii, R$_{arc}$, as naively estimated from 
ground-based imaging as the mean distance between a giant arc and the cluster 
center. Our target selection was not based purely on R$_{arc}$ because our
spectroscopic target list evolved over the course of three semesters (2008A, 2009A, 
and rollover time in 2010A), during which we were actively and continually developing our
complete sample of visually selected giant arcs in the SDSS. Therefore, the clusters
observed in 2008A were selected at a time when we had fewer candidates to choose from
compared to 2009A. Similarly, the list of potential targets in 2010A was larger than
in either 2008A or 2009A. Thus, we tended to select the larger R$_{arc}$ systems, 
but our target clusters are not a subset of our complete giant arc sample with some
simple cut made in R$_{arc}$.

This selection will bias our results in several ways: 1) larger giant arc radii 
will tend to be produced by lensing of higher redshift sources, and 2) clusters 
which produce giant arcs with larger R$_{arc}$ will tend to be 
the most extremely massive systems, even in comparison to typical strong lensing 
selected clusters. Because of this bias we acknowledge that the data presented 
in this paper does not necessarily represent a definitive characterization of 
the ensemble properties of our entire visually selected giant arc sample, nor 
of the cluster lenses which produce those giant arcs. These data do, however, 
serve as the first step in characterizing our complete sample. Our spectroscopic 
follow-up efforts are on-going, and in the future we will target a broader 
range of systems as function of R$_{arc}$. Furthermore, given a large sample of 
strong lensing systems it becomes possible to measure higher order statistics for 
giant arcs, such as the distribution of R$_{arc}$ and the dependence of quantities
such as median source redshift and median lensing cluster mass as a function of
R$_{arc}$. In this context it is not essential that we conduct spectroscopic follow-up
of a random assortment of our giant arc sample, but rather it will be crucial that
we take account for our selection in terms of R$_{arc}$ in future analyses.

Based on modest numbers of spectroscopically confirmed cluster members per 
cluster lens, we have calculated dynamical masses for the foreground lensing 
clusters. The raw masses that we calculate clearly confirm the predictions 
that selecting clusters by strong lensing samples the high-mass tail of the 
mass function at a given epoch of the universe 
\citep[e.g.,][]{dalal2004,hennawi2007,meneghetti2010,fedeli2010}. From our 
sample of $25$ dynamical masses we can compute the median strong lensing cluster 
mass and compare that to the predicted median 
$M_{Vir} = 4.5\times10^{14} M_{\sun} h_{0.7}^{-1}$ 
for strong lensing selected clusters from \citet{hennawi2007}. It is important 
to note that \citet{hennawi2007} calculate the virial mass of their strong lensing 
selected clusters according to the prescription in \citet{bryan1998}, whereas the 
\citet{evrard2008} relation provides a dynamical mass at a fixed 
over-density radius, $R_{200}$. The differences in the subtleties of
how these masses are defined will produce offsets between their values for a 
given cluster halo that can vary as a function of redshift and cosmology 
\citep{hu2003}. To compare our results directly to the median virial mass of 
strong lensing clusters in \citet{hennawi2007} we convert the $M_{200}$ 
values that result from the \citet{evrard2008} scaling relation into 
$M_{Vir}$ values according to the prescription in \citet{hu2003}. We also 
point out that the simulations used in \citet{hennawi2007} were run in a 
cosmology with $\sigma_{8}=0.95$, which is markedly higher than current 
best constraints \citep{komatsu2010}. We can make a simple approximate correction 
for the high $\sigma_{8}$ by simply scaling the \citet{hennawi2007} 
cross-section-weighted median $M_{Vir}$ by the ratio of the mass function 
calculated for $\sigma_{8}=0.95$ and $\sigma_{8}=0.81$, summed over all 
halos with $M_{Vir} > 1\times10^{14} M_{\sun} h_{0.7}^{-1}$, as this is the 
approximate mass where \citet{hennawi2007} find that the cross-section for strong 
lensing becomes negligibly small. It is important to point out that this 
approximation explicitly ignores the effect that $\sigma_{8}$ has on the strong lensing 
cross-section of halos of a given mass, but we assume this to be a sub-dominant 
effect compared to the scaling of the mass function. Taking the fitting formula 
from \citet{jenkins2001} we calculate that the predicted median $M_{Vir}$ 
for $\sigma_{8}=0.81$ should be $\sim7.5\%$ smaller than for the $\sigma_{8}=0.95$ 
used in the simulations in \citet{hennawi2007}, resulting in a predicted 
median $M_{Vir} = 4.16\times10^{14} M_{\sun} h_{0.7}^{-1}$.

The median virial mass of our strong lensing clusters is 
$M_{Vir} = 7.84\times10^{14} M_{\sun} h_{0.7}^{-1}$, approximately $90\%$ larger 
that the prediction from \citet{hennawi2007}. We hesitate to draw strong 
conclusions from the discrepancy in median mass between our cluster lens samples 
and predictions for simulations for several reasons. For one, the errors on our 
dynamical mass estimates are extraordinarily large due to systematic errors associated 
with the small numbers of cluster member redshifts available. We might also be 
concerned with a possible bias in our sample resulting from the selection of lenses with 
larger R$_{arc}$ -- mentioned above -- as targets for Gemini spectroscopy. 
We can examine the data directly for some relationship between R$_{arc}$ and the 
dynamical M$_{200}$ values, and we find no correlation between these two quantities. 
We have reason to expect that this selection is not biasing our median lensing cluster 
mass for the purpose of comparing against \citet{hennawi2007} because in that paper 
the mean virial mass is computed for clusters producing giant arcs with 
$\theta_{arc} > 15\arcsec$, which is comparable to the minimum R$_{arc}$ for our sample 
of spectroscopically observed clusters.

There is however an additional source of predictable bias that should inflate dynamical 
mass estimates of any sample of strong lensing selected clusters. It is understood 
that strong lensing selected clusters as a population are biased with respect to 
several important properties when compared against the general cluster 
population \citep{hennawi2007,oguri2009a,meneghetti2010}. One 
of the notable biases is the spatial orientation of the cluster mass distribution. 
The virialized halos that host galaxy clusters are triaxial, and clusters which 
are efficient strong lenses are more likely to have their major axes aligned along 
the line of sight with respect to the observer, so we must 
assume that our sample of strong lensing selected clusters exhibit have this 
``orientation bias''. We are therefore measuring the projected velocity dispersion of
galaxies that should tend to be preferentially aligned along the major axis 
of the cluster potential. Studies of the position and velocity ellipsoids 
of triaxial halo potentials in N-body simulations find that halo velocity shapes 
are more spherical than halo positional shapes, but that the velocities are still 
significantly triaxial and generally well-aligned with the positional orientation 
of the halo to within $\sim22^{\circ}$ \citep{kasun2005}. This means that the 
projected velocity dispersions measured for a sample of clusters that have an 
orientation bias with the major axis aligned along the line of sight into the sky 
will be biased high with respect to velocity dispersions measured for clusters 
that are randomly oriented on the sky. 

\citet{kasun2005} determine that the average velocity shape for cluster-scale
halos has a minor-major axis ratio of $0.704$ and an intermediate-major axis ratio
of $0.84$. The ratios characterize the relative magnitude of the particle
velocity dispersions in halos projected along the three principle axes
of the halo velocity ellipsoid.  If we were to measure particle velocities -- 
or in real observable terms, member galaxy velocities -- in projection purely 
along the major velocity axis for a sample of clusters, then our resulting 
velocity dispersions would be biased $18\%$ high with respect to the 
average velocity dispersion measured from a sample of randomly oriented clusters. 
Studies of strong lensing halos in simulations find that the population of halos 
that are the most effective strong lenses are not any more triaxial that the general 
halo population \citep{hennawi2007,meneghetti2010}, so we have no reason to expect 
that the positional shapes and velocity shapes of an ensemble of strong lensing clusters 
should have more extreme values for the minor-major and intermediate-major axis 
ratios than the general cluster population. Therefore we take the worst case scenario 
from above for overestimation of the velocity dispersion of a cluster due to 
orientation bias and consider the resulting overestimation of $M_{200}$. We 
use a fit for the virial relation between $\sigma_{v}$ and $M_{200}$ from 
\citet{evrard2008}:
$$M_{200} \propto \sigma_{v}^{\frac{1}{\alpha}}$$   
for which the authors find a best fit $\alpha = 0.3361\pm0.0026$. Given this 
scaling dependence, a sample of measured velocity dispersions that are on 
average $18\%$ high due to orientation bias will result in mass estimates that 
are biased high by $63\%$ on average. This is the extreme case for orientation 
bias, corresponding to a sample of clusters that are all aligned with their 
major axes pointing along the line of sight.

The above computations assume the most extreme possible orientation bias: always 
being aligned with the major axis along the line of sight. Simulations predict that 
strong lensing selected clusters will have a significant orientation bias, but not 
that $all$ strong lensing selected clusters will be perfectly oriented along the line 
of sight. \citet{hennawi2007} predict a median value of $|cos\theta|=0.67$ for 
the alignment angle between the line of sight to the observer and the positional 
major axis for strong lensing selected clusters, compared to the $|cos\theta|=0.5$ 
that you would expect for cluster that are randomly oriented on the sky. 
\citet{meneghetti2010} report predictions for three subsets of strong lensing clusters
defined in different ways: 1) ``critical clusters'' are those which have critical 
lines, 2) clusters which are capable of producing giant arcs, and 3) clusters 
which have a strong lensing cross-section for giant arcs that is larger than
$10^{-3} h^{-2} Mpc^{2}$. We conservatively take the most selective and therefore 
most strongly biased subset -- clusters with strong lensing cross-section for 
giant arcs greater than $10^{-3} h^{-2} Mpc^{2}$ and note that this population 
in the \citet{meneghetti2010} simulations have a median alignment angle of 
$47^{\circ}$, corresponding to $|cos\theta|=0.68$, which is in excellent agreement 
with the results from \citet{hennawi2007}. We combine these two predictions 
for the median alignment angle of the halo major axis and the average axis ratio 
values from \citet{kasun2005} to estimate the average bias we can anticipate in 
velocity dispersions measured for strong lensing selected clusters to be 
$19-20\%$ high relative to dynamical masses measured for a cluster sample 
that is randomly oriented on the sky.

Our estimate of the expected bias in $\sigma_{v}$ measured for strong 
lensing selected clusters assumes that the position and velocity ellipsoids 
for clusters are perfectly aligned, but this turns out not to be the case. 
\citet{kasun2005} measure a median alignment angle of $22^{\circ}$ between 
the position and velocity ellipsoids, where the orientation biases from 
\citet{hennawi2007} and \citet{meneghetti2010} refer to the alignment of 
the position ellipsoid. This tendency toward misalignment should reduce the 
expected bias for velocity dispersions of strong lensing clusters because it 
adds an element of randomization to the orientation of the velocity 
ellipsoid on the sky with respect to the line of sight of the observer. 
This randomness should reduce the impact of the orientation bias 
of strong lensing clusters on velocity dispersion measurements. Specific 
predictions for the magnitude of this reduction require convolving the probability 
distributions for the position orientation angle of strong lensing selected 
clusters from \citet{hennawi2007} and \citet{meneghetti2010} with the probability 
distribution of the orientation angle between the position and velocity principle axes 
from \citet{kasun2005}. The effect should be small, but we do not have the 
necessary probability distributions in hand and leave additional corrections to 
the anticipated dynamical mass bias for future work with higher fidelity data. 
The dynamical mass estimates presented here are intended only to gain a rough 
understanding of $M_{Vir}$ for our cluster sample.

Correcting the predicted median lensing cluster $M_{Vir}$ 
from \citet{hennawi2007} for a $19\%$ bias due to orientation effects we find an 
expected median $M_{Vir} = 5.36\times10^{14} M_{\sun} h_{0.7}^{-1}$, which is 
still $\sim46\%$ small than the median $M_{Vir}$ of our strong lensing cluster sample. 
This kind of discrepancy is not especially problematic when we consider the 
large errors on our dynamical mass estimates. We also note that the semi-analytic models
of \citet{oguri2009a} suggest that the orientation bias for strong lensing clusters
with the largest Einstein radii is likely even more extreme from the predictions
from simulations. Therefore, depending on the true values of the Einstein radii for 
our clusters, it is possible that our sample has a significantly larger underlying 
orientation bias than we accounted for in the preceding calculations, which would 
result in a much larger mass bias. 

We could also be suffering from a selection bias in the sample of cluster member 
galaxies for which we are measuring velocities. Our cluster galaxy redshifts are 
all measured in a field approximately $3\arcmin \times 5\arcmin$ that is centered 
roughly on the cores of our strong lensing clusters, where the size of this field 
is constrained by the field of view of GMOS. We are therefore confining our 
velocity measurements to galaxies that are within the central regions of these 
clusters, with no ability to sample galaxy velocities at larger projected radii 
on the sky. Projected 1D velocities in the cores of clusters should be higher 
than the average projected 1D velocities within $R_{200}$, which is the quantity 
that we use to scale $\sigma_{v}$ into $M_{200}$. Estimating the effect of 
this potential cluster member sampling bias requires knowledge of $R_{200}$ for 
each cluster, and we use equation 8 from \citet{carlberg1997} to estimate $R_{200}$ 
from $\sigma_{v}$ for our cluster lenses. Our clusters have a mean $R_{200}=2.1$ Mpc 
h$^{-1}$, and a mean angular size of the sky of $\theta_{R200}=6\arcmin$. Therefore 
our cluster member sample, which is drawn from within an average angular radius of 
$\sim2.5\arcmin$ of the cluster cores is only sampling cluster galaxies within the 
central $\sim0.42R_{200}$, on average. This sampling bias is likely contributing 
to the high median $M_{Vir}$ that we measure for our cluster lens sample compared 
to the mean $M_{Vir}$ reported in \citet{hennawi2007}.

\section{Conclusions}

We present the results of Gemini/GMOS-North N\&S multi-object spectroscopy 
of $26$ strong lensing selected galaxy clusters. Analysis of our complete 
spectroscopic dataset yields precise redshifts for $69$ likely lensed background 
sources, many of which are multiply imaged by the foreground lensing potentials. 
This dataset dramatically extends the number of strong lensing clusters with 
redshifts available to inform strong lens modeling of the mass structure in the 
cluster cores, especially at $z\gtrsim0.2$. We also characterize the total 
virial masses of our strong lensing clusters via cluster member dynamics for 
comparison against predictions for the typical mass of strong lensing selected 
clusters in simulations. By combining predictions from simulations for the position 
and velocity shapes of halos with predictions for the orientation bias of clusters 
selected by strong lensing we account for the anticipated bias in dynamical masses 
calculated for strong lensing selected clusters, calculating it to be between 
$\sim19-20\%$. The median virial mass of our sample of strong lensing selected 
galaxy clusters is in reasonable agreement with predictions, though still somewhat 
high and possibly suggestive of a more severe orientation bias in our sample 
than is predicted for strong lensing clusters based on simulations.

With the coming era of large area deep imaging surveys (e.g. PanSTARRS, DES, LSST) we 
are poised to extend samples of strong lensing selected galaxy clusters into 
the thousands. In order to take full advantage of future strong lensing cluster 
samples it is crucial that we understand the properties and biases of this 
intriguing subset of galaxy clusters. The analysis presented here is a first step in 
this direction, and we are only beginning to fully exploit the new samples of 
hundreds of strong lensing clusters available in the SDSS and RCS2 surveys. 
Further follow-up of these lens samples will also pave the way for higher order 
analyses, such as combining information from strong lensing with multi-wavelength 
observations (e.g. dynamics of $N\geq50$ cluster members, X-ray, SZ) of a 
$well-selected$ sample of strong lensing clusters in order to quantify the 
biases between different mass observables. These kinds of biases must be 
quantified and thoroughly understood before information gained from analyses 
of strong lensing clusters can be intelligently applied to scaling relations 
and mass estimates for the general cluster population. An empirical 
characterization of strong lensing selected clusters is necessary if we 
hope to take full advantage of the additional information provided by 
strong gravitational lensing in the cores of clusters.

\acknowledgments{We thank the Gemini North observing and support staff for their 
efforts in taking data that contributed to this paper. We give particular thanks 
to Alexander Fritz, Inger J{\o}rgenson, Dara Norman, Knut Olsen, Kathy Roth, and 
Ricardo Shiavon for there help executing our Gemini programs. MBB acknowledges support 
from the Illinois Space Grant Consortium in the form of a graduate fellowship. JFH 
acknowledges support provided by the Alexander von Humboldt Foundation in the framework 
of the Sofja Kovalevskaja Award endowed by the German Federal Ministry of Education 
and Research. The authors also wish to recognize and acknowledge the very significant 
cultural role and reverence that the summit of Mauna Kea has always had within the 
indigenous Hawaiian community. We are most fortunate to have the opportunity to 
conduct observations from this mountain.}

\bibliographystyle{apj}

\begin{thebibliography}{}

\bibitem[Abell et al.(1989)]{abell1989} {Abell}, G.~O., {Corwin}, H.~G.~Jr., {Olowin}, R.~P.~1989, \apjs, 70, 1

\bibitem[Abraham et al.(2004)]{abraham2004} {Abraham}, R.~G. et al.~ 2004, \aj, 127, 2455

\bibitem[Abazajian et al.(2009)]{sdssdr7} {Abazajian}, K.~N. et al. 2009, \apjs, 182, 543

\bibitem[Allen et al.(1992)]{allen1992} {Allen}, S.~W. et al. ~1992, MNRAS, 259, 67

\bibitem[Bayliss et al.(2010)]{bayliss2010} {Bayliss}, M.~B., {Wuyts}, E., {Sharon}, K., {Gladders}, M.~D., {Hennawi}, J.~F., {Koester}, B.~P., {Dahle}, H.~ 2010, \apj, 720, 1559

\bibitem[Bayliss et al.(2011)]{bayliss2011} {Bayliss}, M.~B., {Gladders}, M.~D., {Oguri}, M., {Hennawi}, J.~F., {Sharon}, K., {Koester}, B.~P., {Dahle}, H.~ 2011, \apj, 727, L26

\bibitem[Beers et al.(1990)]{beers1990} {Beers}, T.~C., {Flynn}, K., \& {Gebhardt}, K. 1990, AJ, 100, 32

\bibitem[Belokurov et al.(2009)]{belokurov2009} {Belokurov}, V., {Evans}, N. W., {Hewett}, P. C., {Moiseev}, A., {McMahon}, R. G., {Sanchez}, S. F., {King}, L. J.~2009, MNRAS, 392, 104

\bibitem[B{\"o}hringer et al.(2004)]{bohringer2004} {B{\"o}hringer}, H. et al.~2004, \aap, 425, 367

\bibitem[Bryan \& Norman(1998)]{bryan1998} {Bryan}, G.~L., {Norman}, M.~L. 1998, \apj, 495, 80
 
\bibitem[Burenin et al.(2007)]{burenin2007} {Burenin}, R. A., {Vikhlinin}, A., {Hornstrup}, A., {Ebeling}, H., {Quintana}, H., {Mescheryakov}, A.~2007, \apjs, 172, 561

\bibitem[Carlberg et al.(1997)]{carlberg1997} {Carlberg}, R.~G., {Yee}, H.~K.~C., {Ellingson}, E. ~1997, \apj, 478, 462

\bibitem[Col\'{i}n et al.(2000)]{colin2000} {Col\'{i}n}, P., {Klypin}, A.~A., {Kravtsov}, A.~V.~2000, \apj, 539, 561

\bibitem[Dahle(2006)]{dahle2006} {Dahle}, H.~ 2006, \apj, 653, 954

\bibitem[Dalal et al.(2004)]{dalal2004} {Dalal}, N., {Holder}, G., {Hennawi}, J.~F.~2004, \apj, 609, 50

\bibitem[Diehl et al.(2009)]{diehl2009} {Diehl}, H.~T. et al~ 2009, \apj, 707, 686

\bibitem[Diemand et al.(2004)]{diemand2004} {Diemand}, J., {Moore}, B., {Stadel}, J.~2004, MNRAS, 352, 535

\bibitem[Ebeling et al.(2001)]{macs2001} {Ebeling}, H., {Edge}, A.~C., {Henry}, J.~P.~2001, \apj, 553, 668

\bibitem[Evrard et al.(2008)]{evrard2008} {Evrard}, A.~E. et al.~2008, \apj, 672, 122

\bibitem[Faltenbacher \& Diemand(2006)]{faltenbacher2006} {Faltenbacher}, A. \& {Diemand}, J.~2006, MNRAS, 369, 1698

\bibitem[Faltenbacher et al.(2005)]{faltenbacher2005} {Faltenbacher}, A., {Kravtsov}, A.~V., {Nagai}, D., {Gottl{\"o}ber}, S.~2005, MNRAS, 358, 139

\bibitem[Fedeli et al.(2010)]{fedeli2010} {Fedeli}, C., {Meneghetti}, M., {Gottloeber}, S., {Yepes}, G.~2010, \aap, 519A, 91

\bibitem[Flores et al.(2000)]{flores2000} {Flores},R.~.A., {Maller}, A.~H., {Primack}, J.~R.~2000, \apj, 535, 555

\bibitem[Ghigna et al.(2000)]{ghigna2000} {Ghigna}, S., {Governato}, F. {Lake}, G., {Quinn}, T., {Stadel}, J.~2000, \apj, 544, 616

\bibitem[Gilbank et al.(2011)]{gilbank2011} {Gilbank}, D.~G., {Gladders}, M.~D., {Yee}, H.~K.~C., {Hsieh}, B.~C.~ 2011, \aj, 141, 94

\bibitem[Gladders \& Yee(2005)]{gladders2005} {Gladders}, M.~D. \& {Yee}, H.~K.~C.~ 2005, \apjs, 157, 1

\bibitem[Glazebrook \& Bland-Hawthorn(2001)]{glaze2001} {Glazebrook}, K. \& {Bland-Hawthorn}, J.~ 2001, \pasp, 113, 197

\bibitem[Gunn et al.(1986)]{gunn1986} {Gunn}, J.~E., {Hoessel}, J.~G., {Oke}, J.~B.,~ 1986, \apj, 306, 30

\bibitem[Hennawi et al.(2007)]{hennawi2007} {Hennawi}, J.~F., {Dalal}, N., {Bode}, P.,{Ostriker}, J.~ 2007, \apj, 654, 714

\bibitem[Hennawi et al.(2008)]{hennawi2008} {Hennawi}, J.~F. et al.~ 2008, \aj, 135, 664

\bibitem[Hoekstra \& Jain(2008)]{hoekstra2008} {Hoekstra}, H. \& {Jain}, B.~ 2008, ARNPS, 58, 99

\bibitem[Hook et al.(2004)]{GMOS} {Hook}, I., {J{\o}rgensen}, I., {Allington-Smith}, J.~R., {Davies}, R.~L., {Metcalfe}, N., {Murowinski}, R.~G., {Crampton}, D.~ 2004, \pasp, 116, 425

\bibitem[Horesh et al.(2010)]{horesh2010} {Horesh}, A., {Maoz}, D., {Ebeling}, H., {Seidel}, G., {Bartelmann}, M. 2010, MNRAS, 460, 1318

\bibitem[Hu \& Kravtsov(2003)]{hu2003} {Hu}, W. \& {Kravtsov}, A.~V. 2003, \apj, 584, 702

\bibitem[Jenkins et al.(2001)]{jenkins2001} {Jenkins}, A., {Frenk}, C.~S., {White}, S.~D.~M., {Colberg}, J.~M., {Cole}, S., {Evrard}, A.~E., {Couchman}, H.~M.~P., {Yoshida}, N. 2001, MNRAS, 321, 372

\bibitem[Kasun \& Evrard(2005)]{kasun2005} {Kasun}, S.~F., {Evrard}, A.~E. 2005, \apj, 629, 781

\bibitem[Koester et al.(2007)]{koester2007} {Koester}, B.~P. et al.~2007, \apj, 660, 23

\bibitem[Koester et al.(2010)]{koester2010} {Koester}, B.~P., {Gladders}, M.~D., {Hennawi}, J.~F., {Sharon}, K., {Wuyts}, E., {Rigby}, J.~R., {Bayliss}, M.~B., {Dahle}, H.~ 2010, \apj, 723L, 73

\bibitem[Komatsu et al.(2010)]{komatsu2010} {Komatsu}, et al.~2010, ArXiv, e-prints, astro-ph/1001.4538

\bibitem[Kubo et al.(2009)]{kubo2009} {Kubo}, J.~M., {Allam}, S.~S., {Annis}, J., {Buckley-Geer}, E.~J., {Diehl}, H.~T., {Kubik}, D., {Lin}, H. {Tucker}, D.~ 2009, \apj, 696L, 61

\bibitem[Limousin et al.(2008)]{limousin2008} {Limousin}, M. et al.~2008, A\&A, 489, 23L

\bibitem[Meneghetti et al.(2010)]{meneghetti2010} {Meneghetti}, M., {Fedeli}, C., {Pace}, F., {Gottloeber}, S., {Yepes}, G.,~2010, \aap, 519A, 90

\bibitem[Ofek et al.(2008)]{ofek2008} {Ofek}, E.~O., {Seitz}, S., {Klein}, F.~2008, MNRAS, 389, 311

\bibitem[Oguri \& Blandford(2009)]{oguri2009a} {Oguri}, M. \& Blandford, R.~D.~ 2009, MNRAS, 392, 930

\bibitem[Oguri et al.(2009)]{oguri2009b} {Oguri}, M. et al.~2009, \apj, 699, 1038

\bibitem[Okabe et al.(2010)]{okabe2009} {Okabe}, N., {Takada}, M. ,{Umetsu}, K., {Futamase}, T., {Smith}, G.~P.,~2010 \pasj, 62, 811

\bibitem[Richard et al.(2009)]{richard2009} {Richard}, J., {Pei}, L., {Limousin}, M., {Jullo}, E., {Kneib}, J.~P.~2009, \aap, 498, 37

\bibitem[Richard et al.(2010)]{richard2010} {Richard}, J. et al.~2010, MNRAS, 404, 325

\bibitem[Rizza et al.(2003)]{rizza2003} {Rizza}, E., {Morrison}, G.~E., {Owen}, F.~N., {Ledlow}, M.~J., {Burns}, J.~O., {Hill}, J. ~2003, \aj, 126, 119

\bibitem[Rozo et al.(2009a)]{rozo2009a} {Rozo}, E. et al. 2009a, \apj, 699, 768

\bibitem[Rozo et al.(2009b)]{rozo2009b} {Rozo}, E. et al. 2009b, \apj, 703, 601

\bibitem[Shapley et al.(2003)]{shapley2003} {Shapley}, A., {Steidel}, C., {Pettini}, M., {Adelberger}, K.~L.~ 2003, \apj, 588, 65

\bibitem[Sheldon et al.(2009)]{sheldon2009} {Sheldon}, E.~S. et al. 2009, \apj, 703, 2217

\bibitem[Smail et al.(2002)]{smail2002} Smail, I., Ivison, R.~J., Blain, A.~W., Kneib, J.-P. ~2002, MNRAS, 331, 495

\bibitem[Smith (1936)]{smith1936} {Smith}, S.~ 1936, \apj, 83, 23S

\bibitem[Vanderlinde et al.(2010)]{vanderlinde2010} {Vanderlinde}, K. et al.~ 2010, \apj, 722, 1180

\bibitem[Vikhlinin et al.(2009a)]{vikhlinin2009a} {Vikhlinin}, A. et al.~ 2009a, \apj, 692, 1033

\bibitem[Vikhlinin et al.(2009b)]{vikhlinin2009b} {Vikhlinin}, A. et al.~ 2009b, \apj, 692, 1060

\bibitem[Voges et al.(1999)]{voges1999} {Voges} W. et al.~1999, \aap, 349, 389

\bibitem[Wen et al.(2009)]{wen2009} {Wen} Z., et al. 2009~ Research in Astron. Astrophys., 9, 5

\bibitem[White et al.(2010)]{white2010} {White}, M., {Cohn}, J.~D., {Smit}, R.~ 2010, MNRAS, 408, 1818

\bibitem[Wuyts et al.(2010)]{wuyts2010} {Wuyts}, E. et al.~2010, \apj, 724, 1182

\bibitem[York et al.(2000)]{york2000} {York}, Donald G. et al~ 2000, \aj, 120, 1579

\bibitem[Zwicky(1937)]{zwicky1937} {Zwicky}, F. 1937, \apj, 86, 217

\end{thebibliography}

\clearpage

\LongTables
\begin{deluxetable*}{cccccc}
\tablecaption{Individual Redshifts Measured with Gemini/GMOS\label{tabredshifts}}
\tablewidth{0pt}
\tabletypesize{\tiny}
\tablehead{
\colhead{Cluster Core} &
\colhead{Label} &
\colhead{$\alpha$\tablenotemark{a}} &
\colhead{$\delta$\tablenotemark{a}} &
\colhead{$z$} &
\colhead{Redshift Class\tablenotemark{b}} }
\startdata
SDSS J0851+3331  &  --  &   --  &  --  &  --  &  --  \\
--  &  A1  &  08:51:37.108  &  +33:31:13.512  &  1.6926  &  3   \\
--  &  A2  &  08:51:37.115  &  +33:31:06.522  &  1.6924  &  3   \\
--  &  A3  &  08:51:39.363  &  +33:31:27.012  &  1.6925  & 2   \\
--  &  B1  &  08:51:37.987  &  +33:31:06.879  &  1.346  &  3   \\
--  &  B2  &  08:51:38.021  &  +33:31:03.185  &  1.346  &  3   \\
--  &  C2  &  08:51:40.139  &  +33:31:21.065  &  1.2539  &  3   \\
--  &  gal  &  08:51:44.675  &  +33:29:46.418  &  0.3712  &  3  \\
--  &  gal  &  08:51:39.923  &  +33:30:48.656  &  0.3744  &  3  \\
--  &  gal  &  08:51:38.584  &  +33:29:54.411  &  0.3729  &  3  \\
--  &  gal  &  08:51:42.254  &  +33:30:52.391  &  0.3709  &  3  \\
--  &  gal  &  08:51:37.794  &  +33:30:00.851  &  0.3700  &  3  \\
--  &  gal  &  08:51:46.587  &  +33:30:00.618  &  0.3689  &  3  \\
--  &  gal  &  08:51:35.226  &  +33:31:00.823  &  0.3693  &  3  \\
--  &  gal  &  08:51:37.908  &  +33:31:29.772  &  0.3654  &  3  \\
--  &  gal  &  08:51:31.927  &  +33:30:48.615  &  0.3709  &  3  \\
--  &  gal  &  08:51:44.623  &  +33:30:47.351  &  0.3689  &  3  \\
--  &  gal  &  08:51:29.479  &  +33:31:49.341  &  0.3797  &  3  \\
--  &  gal  &  08:51:36.775  &  +33:31:55.549  &  0.3695  &  3  \\
--  &  gal  &  08:51:37.266  &  +33:31:01.153  &  0.3590  &  3  \\
--  &  gal  &  08:51:36.061  &  +33:31:10.532  &  0.3722  &  3  \\
--  &  gal  &  08:51:38.738  &  +33:31:17.344  &  0.3698  &  3  \\
SDSS J0915+3826  &  --  &   --  &  --  &  --  &  --  \\
--  &  A1  &  09:15:38.147  &  +38:27:04.617  &  1.501  &  3   \\
--  &  A2  &  09:15:37.999  &  +38:26:57.929  &  1.501  &  3   \\
--  &  A3  &  09:15:38.531  &  +38:27:10.096  &  1.501  &  3   \\
--  &  B1  &  09:15:40.948  &  +38:26:52.628  &  5.200  &  3   \\
--  &  C1  &  09:15:43.090  &  +38:27:05.455  &  1.436  &  2   \\
--  &  gal  &  09:15:47.306  &  +38:26:51.049  &  0.3966  &  3  \\
--  &  gal  &  09:15:35.280  &  +38:25:51.613  &  0.3986  &  3  \\
--  &  gal  &  09:15:34.803  &  +38:26:02.915  &  0.3979  &  3  \\
--  &  gal  &  09:15:31.108  &  +38:28:11.743  &  0.3932  &  3  \\
--  &  gal  &  09:15:38.085  &  +38:25:59.715  &  0.4067  &  3  \\
--  &  gal  &  09:15:43.870  &  +38:26:38.909  &  0.3985  &  3  \\
--  &  gal  &  09:15:38.438  &  +38:27:08.188  &  0.4026  &  3  \\
--  &  gal  &  09:15:37.848  &  +38:27:20.108  &  0.3937  &  3  \\
--  &  gal  &  09:15:28.819  &  +38:27:05.551  &  0.3994  &  2  \\
--  &  gal  &  09:15:44.591  &  +38:26:48.453  &  0.3985  &  3  \\
--  &  gal  &  09:15:49.957  &  +38:28:24.680  &  0.3952  &  3  \\
--  &  gal  &  09:15:41.686  &  +38:26:34.089  &  0.3940  &  3  \\
--  &  gal  &  09:15:39.709  &  +38:26:55.691  &  0.3979  &  3  \\
--  &  gal  &  09:15:39.928  &  +38:27:08.188  &  0.3992  &  3  \\
--  &  gal  &  09:15:39.482  &  +38:27:14.518  &  0.3892  &  3  \\
--  &  gal  &  09:15:42.404  &  +38:27:02.392  &  0.3920  &  3  \\
SDSS J0957+0509  &  --  &   --  &  --  &  --  &  --  \\
--  &  A1  &  09:57:38.826  &  +05:09:25.092  &  1.821  &  3   \\
--  &  A2  &  09:57:38.709  &  +05:09:28.189  &  1.821  &  3   \\
--  &  A3  &  09:57:38.627  &  +05:09:31.392  &  1.820  &  3   \\
--  &  B1  &  09:57:41.692  &  +05:09:38.161  &  1.007  &  2   \\
--  &  C1  &  09:57:37.961  &  +05:09:14.796  &  1.926  &  2   \\
--  &  gal  &  09:57:39.262  &  +05:07:18.083  &  0.4499  &  3  \\
--  &  gal  &  09:57:48.226  &  +05:10:46.056  &  0.4436  &  3  \\
--  &  gal  &  09:57:39.883  &  +05:09:31.320  &  0.4375  &  3  \\
--  &  gal  &  09:57:40.154  &  +05:09:40.068  &  0.4433  &  3  \\
--  &  gal  &  09:57:40.130  &  +05:09:48.201  &  0.4503  &  3  \\
--  &  gal  &  09:57:40.220  &  +05:09:56.376  &  0.4495  &  3  \\
--  &  gal  &  09:57:43.042  &  +05:11:28.860  &  0.4516  &  2  \\
--  &  gal  &  09:57:40.429  &  +05:09:17.568  &  0.4517  &  3  \\
SDSS J1028+1324  &  --  &  --  &  --  &  --  &  --  \\
--  &  A1  &  10:28:04.503  &  +13:25:12.474  &  ...  &  0   \\
--  &  B1  &  10:28:04.966  &  +13:25:09.415  &  ...  &  0   \\
--  &  C1  &  10:28:05.117  &  +13:25:02.933  &  ...  &  0   \\
--  &  D1  &  10:28:03.624  &  +13:24:52.493  &  ...  &  0   \\
--  &  gal  &  10:28:15.321  &  +13:25:41.619  &  0.4181  &  3  \\
--  &  gal  &  10:28:06.316  &  +13:23:50.190  &  0.4165  &  3  \\
--  &  gal  &  10:28:06.559  &  +13:24:39.220  &  0.4177  &  3  \\
--  &  gal  &  10:28:01.451  &  +13:25:03.410  &  0.4110  &  3  \\
--  &  gal  &  10:28:01.845  &  +13:25:22.561  &  0.4125  &  3  \\
--  &  gal  &  10:28:05.011  &  +13:26:13.956  &  0.4138  &  3  \\
--  &  gal  &  10:28:04.129  &  +13:24:53.018  &  0.4134  &  3  \\
--  &  gal  &  10:28:04.637  &  +13:25:11.420  &  0.4160  &  2  \\
--  &  gal  &  10:28:06.185  &  +13:25:47.510  &  0.4129  &  3  \\
SDSS J1038+4849  &  --  &   --  &  --  &  --  &  --  \\
--  &  A1  &  10:38:42.465  &  +48:49:30.154  &  2.198  &  3   \\
--  &  A2  &  10:38:41.772  &  +48:49:18.893  &  2.198  &  3   \\
--  &  B1  &  10:38:43.461  &  +48:49:10.063  &  0.9660  &  3   \\
--  &  B2  &  10:38:44.062  &  +48:49:17.767  &  0.9652  &  3   \\
--  &  C1  &  10:38:42.613  &  +48:49:11.903  &  2.783  &  1   \\
--  &  C2  &  10:38:42.362  &  +48:49:14.279  &  2.783  &  1   \\
--  &  D1  &  10:38:42.743  &  +48:49:05.462  &  0.8020  &  3   \\
--  &  gal  &  10:38:40.941  &  +48:49:35.262  &  0.4309  &  3  \\
--  &  gal  &  10:38:44.068  &  +48:49:06.107  &  0.4337  &  2  \\
--  &  gal  &  10:38:43.200  &  +48:49:37.391  &  0.4310  &  3  \\
--  &  gal  &  10:38:49.469  &  +48:48:51.235  &  0.4265  &  3  \\
--  &  gal  &  10:38:48.370  &  +48:47:46.292  &  0.4307  &  3  \\
--  &  gal  &  10:38:54.416  &  +48:50:40.137  &  0.4338  &  3  \\
RCS2 J1055+5548  &  --  &   --  &  --  &  --  &  --  \\
--  &  A1  &  10:55:03.791  &  +55:48:09.597  &  1.2499  &  3   \\
--  &  B1  &  10:55:05.350  &  +55:48:10.640  &  0.9358  &  3   \\
--  &  B2  &  10:55:04.653  &  +55:48:09.638  &  0.9360  &  3   \\
--  &  C1  &  10:55:06.469  &  +55:48:31.665  &  0.7769  &  3   \\
--  &  D1  &  10:55:02.813  &  +55:48:36.568  &  ...   &  0   \\
--  &  gal  &  10:55:02.566  &  +55:48:56.151  &  0.4704  &  3  \\
--  &  gal  &  10:54:55.593  &  +55:48:34.041  &  0.4668  &  3  \\
--  &  gal  &  10:54:56.922  &  +55:48:55.423  &  0.4651  &  3  \\
--  &  gal  &  10:55:04.468  &  +55:48:16.875  &  0.4628  &  3  \\
--  &  gal  &  10:55:04.245  &  +55:48:02.002  &  0.4590  &  3  \\
--  &  gal  &  10:55:05.937  &  +55:48:44.698  &  0.4654  &  3  \\
--  &  gal  &  10:55:07.259  &  +55:48:00.547  &  0.4644  &  2  \\
--  &  gal  &  10:55:02.391  &  +55:48:15.790  &  0.4676  &  3  \\
--  &  gal  &  10:55:09.861  &  +55:48:27.669  &  0.4682  &  3  \\
--  &  gal  &  10:55:03.994  &  +55:48:35.085  &  0.4656  &  3  \\
--  &  gal  &  10:55:09.003  &  +55:49:31.568  &  0.4705  &  2  \\
SDSS J1115+5319  &  --  &   --  &  --  &  --  &  --  \\
--  &  A1  &  11:15:16.352  &  +53:19:22.807  &  ...  &  0   \\
--  &  A2  &  11:15:16.624  &  +53:19:23.919  &  ...  &  0   \\
--  &  B1  &  11:15:17.994  &  +53:19:05.888  &  ...  &  0   \\
--  &  C1  &  11:15:18.399  &  +53:19:51.179  &  ...  &  0   \\
--  &  D1  &  11:15:13.850  &  +53:19:37.171  &  1.234  &  2   \\
--  &  E1  &  11:15:17.994  &  +53:19:05.888  &  ...  &  0   \\
--  &  gal  &  11:15:20.833  &  +53:21:01.052  &  0.4601  &  3  \\
--  &  gal  &  11:15:19.714  &  +53:18:37.955  &  0.4670  &  3  \\
--  &  gal  &  11:15:12.720  &  +53:19:30.758  &  0.4586  &  3  \\
--  &  gal  &  11:15:12.263  &  +53:18:30.924  &  0.4745  &  3  \\
--  &  gal  &  11:15:17.849  &  +53:19:49.449  &  0.4640  &  3  \\
--  &  gal  &  11:15:17.314  &  +53:21:15.266  &  0.4639  &  2  \\
--  &  gal  &  11:15:15.106  &  +53:20:02.907  &  0.4708  &  3  \\
--  &  gal  &  11:15:14.519  &  +53:18:53.281  &  0.4660  &  3  \\
--  &  gal  &  11:15:14.485  &  +53:19:48.831  &  0.4654  &  3  \\
--  &  gal  &  11:15:07.378  &  +53:19:55.738  &  0.4642  &  3  \\
--  &  gal  &  11:15:05.538  &  +53:20:42.403  &  0.4654  &  3  \\
--  &  gal  &  11:15:04.216  &  +53:21:00.448  &  0.4671  &  3  \\
--  &  gal  &  11:15:10.107  &  +53:19:39.945  &  0.4759  &  3  \\
--  &  gal  &  11:15:09.792  &  +53:19:25.251  &  0.4702  &  3  \\
SDSS J1138+2754  &  --  &   --  &  --  &  --  &  --  \\
--  &  A1  &  11:38:09.499  &  +27:54:45.152  &  1.3338  &  3   \\
--  &  A2  &  11:38:08.717  &  +27:54:44.651  &  1.3335  &  3   \\
--  &  A3  &  11:38:07.937  &  +27:54:38.602  &  1.3332  &  3   \\
--  &  B1  &  11:38:08.909  &  +27:54:39.110  &  0.9094  &  3   \\
--  &  B2  &  11:38:08.050  &  +27:54:37.428  &  0.9092  &  2   \\
--  &  B3  &  11:38:08.318  &  +27:54:36.329  &  0.9091  &  3   \\
--  &  C1  &  11:38:08.830  &  +27:54:51.126  &  1.455   &  1   \\
--  &  D1  &  11:38:09.839  &  +27:54:11.212  &  ...   &  0  \\
--  &  D2  &  11:38:10.138  &  +27:54:12.976  &  ...   &  0  \\
--  &  gal  &  11:38:11.892  &  +27:53:37.779  &  0.4593  &  3  \\
--  &  gal  &  11:38:12.263  &  +27:55:50.191  &  0.4660  &  3  \\
--  &  gal  &  11:38:07.045  &  +27:56:09.891  &  0.4478  &  3  \\
--  &  gal  &  11:38:11.861  &  +27:55:17.109  &  0.4646  &  3  \\
--  &  gal  &  11:38:10.107  &  +27:53:25.907  &  0.4495  &  3  \\
--  &  gal  &  11:38:08.854  &  +27:54:01.997  &  0.4489  &  2  \\
--  &  gal  &  11:38:09.949  &  +27:52:51.190  &  0.4431  &  2  \\
--  &  gal  &  11:38:08.318  &  +27:55:51.819  &  0.4510  &  3  \\
--  &  gal  &  11:38:10.303  &  +27:54:24.608  &  0.4485  &  3  \\
--  &  gal  &  11:38:08.727  &  +27:54:37.805  &  0.4544  &  2  \\
--  &  gal  &  11:38:04.851  &  +27:55:42.316  &  0.4431  &  3  \\
SDSS J1152+3313  &  --  &   --  &  --  &  --  &  --  \\
--  &  A1  &  11:51:59.671  &  +33:13:38.358  &  2.491  &  1   \\
--  &  A2  &  11:52:00.028  &  +33:13:34.540  &  2.491  &  1   \\
--  &  B1  &  11:52:01.003  &  +33:13:47.902  &  4.1422  &  3   \\
--  &  B2  &  11:52:00.838  &  +33:13:33.359  &  4.1423   &  3   \\
--  &  gal  &  11:52:04.323  &  +33:12:56.816  &  0.3650  &  3  \\
--  &  gal  &  11:52:04.340  &  +33:12:04.438  &  0.3586  &  3  \\
--  &  gal  &  11:52:00.866  &  +33:12:59.150  &  0.3573  &  3  \\
--  &  gal  &  11:51:59.025  &  +33:12:01.198  &  0.3581  &  3  \\
--  &  gal  &  11:51:53.402  &  +33:12:03.642  &  0.3627  &  3  \\
--  &  gal  &  11:52:00.227  &  +33:13:43.727  &  0.3631  &  3  \\
--  &  gal  &  11:52:08.268  &  +33:14:06.181  &  0.3559  &  3  \\
--  &  gal  &  11:51:58.634  &  +33:13:46.488  &  0.3670  &  3  \\
--  &  gal  &  11:52:02.019  &  +33:12:58.752  &  0.3601  &  3  \\
--  &  gal  &  11:52:00.052  &  +33:13:57.694  &  0.3655  &  3  \\
--  &  gal  &  11:52:01.120  &  +33:14:04.670  &  0.3630  &  3  \\
--  &  gal  &  11:51:55.510  &  +33:12:57.914  &  0.3641  &  3  \\
--  &  gal  &  11:52:01.666  &  +33:14:14.036  &  0.3667  &  3  \\
--  &  gal  &  11:52:01.010  &  +33:14:15.478  &  0.3597  &  3  \\
SDSS J1152+0930  &  --  &   --  &  --  &  --  &  --  \\
--  &  A1  &  11:52:48.024  &  +09:30:08.583  &  0.8930  &  3   \\
--  &  A2  &  11:52:46.916  &  +09:30:14.196  &  0.8945  &  3   \\
--  &  A3  &  11:52:46.840  &  +09:30:06.101  &  0.8932  &  2   \\
--  &  B1  &  11:52:47.506  &  +09:30:41.844  &  0.9760  &  1   \\
--  &  C1  &  11:52:47.873  &  +09:30:06.636  &  ...  &  0   \\
--  &  D1  &  11:52:47.413  &  +09:30:29.605  &  ...  &  0   \\
--  &  D2  &  11:52:46.648  &  +09:30:23.198  &  ...  &  0   \\
--  &  E1  &  11:52:46.376  &  +09:30:17.077  &  ...  &  0   \\
--  &  gal  &  11:52:46.442  &  +09:31:10.913  &  0.5078  &  3  \\
--  &  gal  &  11:52:45.765  &  +09:29:50.322  &  0.5242  &  3  \\
--  &  gal  &  11:52:49.240  &  +09:28:47.061  &  0.5178  &  3  \\
--  &  gal  &  11:52:45.178  &  +09:31:11.761  &  0.5069  &  3  \\
--  &  gal  &  11:52:48.117  &  +09:29:32.407  &  0.5212  &  3  \\
SDSS J1209+2640  &  --  &   --  &  --  &  --  &  --  \\
--  &  A1  &  12:09:24.344  &  +26:40:52.444  &  1.021  &  3   \\
--  &  B1  &  12:09:23.963  &  +26:40:50.178  &  0.879  &  1   \\
--  &  C1  &  12:09:21.879  &  +26:40:56.007  &  3.948  &  2   \\
--  &  C2  &  12:09:22.273  &  +26:41:04.934  &  3.948  &  2   \\
--  &  D1  &  12:09:22.016  &  +26:40:44.994  &  ...  &  0   \\
--  &  E1  &  12:09:21.223  &  +26:40:46.971  &  ...  &  0   \\
--  &  F1  &  12:09:24.955  &  +26:40:52.313  &  ...  &  0   \\
--  &  gal  &  12:09:26.510  &  +26:40:21.895  &  0.5760  &  3  \\
--  &  gal  &  12:09:20.941  &  +26:40:18.475  &  0.5684  &  3  \\
--  &  gal  &  12:09:23.015  &  +26:40:30.299  &  0.5576  &  3  \\
--  &  gal  &  12:09:21.542  &  +26:40:30.231  &  0.5620  &  3  \\
--  &  gal  &  12:09:24.687  &  +26:39:47.473  &  0.5654  &  3  \\
--  &  gal  &  12:09:23.393  &  +26:40:44.073  &  0.5586  &  3  \\
--  &  gal  &  12:09:22.349  &  +26:40:50.473  &  0.5667  &  3  \\
--  &  gal  &  12:09:18.054  &  +26:40:55.355  &  0.5534  &  3  \\
--  &  gal  &  12:09:27.063  &  +26:41:17.671  &  0.5564  &  3  \\
--  &  gal  &  12:09:20.972  &  +26:40:24.566  &  0.5545  &  3  \\
--  &  gal  &  12:09:23.015  &  +26:40:30.299  &  0.5575  &  3  \\
--  &  gal  &  12:09:18.905  &  +26:41:23.494  &  0.5665  &  3  \\
--  &  gal  &  12:09:23.053  &  +26:40:37.935  &  0.5525  &  3  \\
--  &  gal  &  12:09:18.552  &  +26:41:30.065  &  0.5641  &  3  \\
--  &  gal  &  12:09:18.833  &  +26:41:01.590  &  0.5553  &  3  \\
SDSS J1226+2152  &  --  &   --  &  --  &  --  &  --  \\
--  &  A1  &  12:26:51.691  &  +21:52:14.489  &  2.9233  &  2   \\
--  &  A2  &  12:26:51.375  &  +21:52:14.310  &  2.9233  &  2   \\
--  &  B1  &  12:26:51.962  &  +21:52:34.978  &  1.3358  &  2   \\
--  &  C1  &  12:26:54.341  &  +21:52:23.134  &  0.7278  &  2   \\
--  &  D1  &  12:26:51.313  &  +21:52:17.325  &  0.7718  &  3   \\
--  &  E1  &  12:26:52.079  &  +21:52:24.424  &  0.7323  &  3   \\
--  &  gal  &  12:26:48.155  &  +21:52:53.415  &  0.4304  &  3  \\
--  &  gal  &  12:26:48.453  &  +21:53:22.110  &  0.4328  &  3  \\
--  &  gal  &  12:26:49.322  &  +21:53:19.844  &  0.4315  &  3  \\
--  &  gal  &  12:26:51.183  &  +21:51:11.358  &  0.4330  &  3  \\
--  &  gal  &  12:26:50.489  &  +21:52:53.991  &  0.4407  &  3  \\
--  &  gal  &  12:26:50.991  &  +21:52:26.635  &  0.4321  &  3  \\
--  &  gal  &  12:26:51.749  &  +21:52:24.974  &  0.4375  &  3  \\
--  &  gal  &  12:26:52.312  &  +21:51:44.874  &  0.4388  &  3  \\
--  &  gal  &  12:26:52.920  &  +21:52:31.456  &  0.4338  &  3  \\
--  &  gal  &  12:26:54.595  &  +21:52:35.672  &  0.4384  &  3  \\
--  &  gal  &  12:26:50.915  &  +21:52:28.064  &  0.4375  &  3  \\
SDSS J1226+2149  &  --  &   --  &  --  &  --  &  --  \\
--  &  A1  &  12:26:50.153  &  +21:50:07.768  &  1.6045  &  3   \\
--  &  A2  &  12:26:50.445  &  +21:50:10.432  &  1.6045  &  3   \\
--  &  B1  &  12:26:52.014  &  +21:49:57.084  &  0.8011  &  3   \\
--  &  B2  &  12:26:51.924  &  +21:50:00.105  &  0.8014  &  3   \\
--  &  C1  &  12:26:52.439  &  +21:50:14.614  &  0.9134  &  2   \\
--  &  D1  &  12:26:51.564  &  +21:50:17.828  &  1.1353  &  3   \\
--  &  E1  &  12:26:51.506  &  +21:49:32.385  &  ...  &  0   \\
--  &  F1  &  12:26:52.086  &  +21:49:31.122  &  ...  &  0   \\
--  &  gal  &  12:26:50.088  &  +21:50:29.734  &  0.4395  &  3  \\
--  &  gal  &  12:26:51.523  &  +21:48:55.265  &  0.4399  &  3  \\
--  &  gal  &  12:26:50.939  &  +21:49:55.745  &  0.4385  &  3  \\
--  &  gal  &  12:26:51.736  &  +21:49:42.568  &  0.4326  &  3  \\
--  &  gal  &  12:26:52.278  &  +21:49:52.003  &  0.4348  &  3  \\
--  &  gal  &  12:26:48.069  &  +21:50:22.325  &  0.4339  &  3  \\
--  &  gal  &  12:26:52.261  &  +21:50:47.298  &  0.4362  &  3  \\
--  &  gal  &  12:26:49.078  &  +21:49:59.529  &  0.4310  &  3  \\
--  &  gal  &  12:26:49.535  &  +21:50:06.519  &  0.4362  &  3  \\
--  &  gal  &  12:26:57.569  &  +21:49:49.661  &  0.4336  &  3  \\
--  &  gal  &  12:26:54.410  &  +21:49:00.264  &  0.4352  &  3  \\
Abell 1703  &  --  &   --  &  --  &  --  &  --  \\
--  &  A1  &  13:15:06.492  &  +51:49:04.048  &  0.889  &  3   \\
--  &  A2  &  13:15:06.643  &  +51:49:07.083  &  0.889  &  3   \\
--  &  A3  &  13:15:05.826  &  +51:49:04.803  &  0.889  &  3   \\
--  &  A4  &  13:15:06.046  &  +51:49:10.887  &  0.889  &  3   \\
--  &  gal  &  13:15:11.072  &  +51:46:53.722  &  0.2690  &  3  \\
--  &  gal  &  13:15:02.262  &  +51:49:51.179  &  0.2707  &  3  \\
--  &  gal  &  13:14:58.098  &  +51:49:16.256  &  0.2886  &  3  \\
GHO 132029+315500  &  --  &   --  &  --  &  --  &  --  \\
--  &  A1  &  13:22:50.404  &  +31:39:15.084  &  ...  &  0   \\
--  &  A2  &  13:22:49.621  &  +31:39:00.253  &  ...  &  0   \\
--  &  A3  &  13:22:50.424  &  +31:39:21.703  &  ...  &  0   \\
--  &  B1  &  13:22:46.713  &  +31:39:33.260  &  0.8473  &  3   \\
--  &  C1  &  13:22:46.167  &  +31:38:55.892  &  1.1513  &  3   \\
--  &  D1  &  13:22:47.616  &  +31:39:29.984  &  0.8121  &  3   \\
--  &  gal  &  13:22:49.391  &  +31:39:31.392  &  0.3070  &  3  \\
--  &  gal  &  13:22:46.864  &  +31:39:43.340  &  0.3069  &  3  \\
--  &  gal  &  13:22:48.907  &  +31:38:55.460  &  0.3072  &  3  \\
--  &  gal  &  13:22:44.032  &  +31:39:20.447  &  0.3194  &  3  \\
--  &  gal  &  13:22:55.684  &  +31:38:50.241  &  0.3126  &  3  \\
--  &  gal  &  13:22:44.677  &  +31:38:53.270  &  0.3089  &  3  \\
--  &  gal  &  13:22:54.661  &  +31:37:33.488  &  0.3095  &  3  \\
--  &  gal  &  13:22:48.406  &  +31:38:15.614  &  0.3038  &  3  \\
--  &  gal  &  13:22:49.099  &  +31:38:48.147  &  0.2996  &  3  \\
RXC J1327.0+0211  &  --  &   --  &  --  &  --  &  --  \\
--  &  A1  &  13:27:06.969  &  +02:12:47.633  &  0.990  &  3   \\
--  &  A2  &  13:27:06.825  &  +02:12:51.812  &  0.990  &  3   \\
--  &  B1  &  13:27:03.443  &  +02:12:20.074  &  1.4760  &  3   \\
--  &  C1  &  13:27:03.491  &  +02:12:05.128  &  1.602   &  2   \\
--  &  gal  &  13:27:04.634  &  +02:10:54.027  &  0.2608  &  2  \\
--  &  gal  &  13:27:09.362  &  +02:11:45.615  &  0.2627  &  3  \\
--  &  gal  &  13:26:59.780  &  +02:11:24.444  &  0.2582  &  3  \\
--  &  gal  &  13:26:59.213  &  +02:10:19.899  &  0.2600  &  3  \\
--  &  gal  &  13:27:10.797  &  +02:12:37.348  &  0.2526  &  3  \\
--  &  gal  &  13:27:01.857  &  +02:12:17.907  &  0.2570  &  3  \\
--  &  gal  &  13:27:08.291  &  +02:14:29.102  &  0.2591  &  2  \\
SDSS J1343+4155  &  --  &   --  &  --  &  --  &  --  \\
--  &  A1  &  13:43:33.853  &  +41:55:08.917  &  2.091  &  3   \\
--  &  B1  &  13:43:30.691  &  +41:54:55.212  &  4.994  &  2   \\
--  &  C1  &  13:43:35.110  &  +41:54:55.418  &  1.2936  &  3   \\
--  &  D1  &  13:43:32.442  &  +41:54:48.826  &  0.9516  &  3   \\
--  &  gal  &  13:43:26.640  &  +41:53:01.009  &  0.4207  &  3  \\
--  &  gal  &  13:43:33.184  &  +41:53:46.039  &  0.4113  &  3  \\
--  &  gal  &  13:43:31.584  &  +41:54:42.523  &  0.4199  &  3  \\
--  &  gal  &  13:43:34.997  &  +41:55:34.804  &  0.4189  &  3  \\
--  &  gal  &  13:43:33.445  &  +41:55:54.813  &  0.4270  &  3  \\
SDSS J1420+3955  &  --  &   --  &  --  &  --  &  --  \\
--  &  A1  &  14:20:38.544  &  +39:54:53.825  &  2.161  &  2   \\
--  &  B1  &  14:20:37.445  &  +39:54:49.471  &  3.0665  &  3   \\
--  &  B2  &  14:20:37.689  &  +39:54:45.901  &  3.0665  &  3   \\
--  &  gal  &  14:20:37.205  &  +39:55:23.543  &  0.6134  &  3  \\
--  &  gal  &  14:20:37.442  &  +39:55:25.877  &  0.6044  &  3  \\
--  &  gal  &  14:20:38.513  &  +39:55:44.746  &  0.6046  &  3  \\
--  &  gal  &  14:20:38.695  &  +39:54:52.053  &  0.6158  &  3  \\
--  &  gal  &  14:20:40.353  &  +39:54:43.003  &  0.6082  &  3  \\
--  &  gal  &  14:20:41.150  &  +39:55:02.518  &  0.6009  &  2  \\
--  &  gal  &  14:20:41.304  &  +39:54:56.036  &  0.6143  &  3  \\
--  &  gal  &  14:20:42.540  &  +39:55:17.528  &  0.6003  &  3  \\
--  &  gal  &  14:20:42.475  &  +39:54:48.950  &  0.5983  &  2  \\
--  &  gal  &  14:20:43.309  &  +39:55:08.134  &  0.6112  &  3  \\
--  &  gal  &  14:20:39.605  &  +39:55:38.951  &  0.6102  &  3  \\
--  &  gal  &  14:20:42.255  &  +39:54:13.230  &  0.6120  &  3  \\
--  &  gal  &  14:20:40.446  &  +39:55:10.043  &  0.6026  &  3  \\
SDSS J1446+3033  &  --  &   --  &  --  &  --  &  --  \\
--  &  A1  &  14:46:29.930  &  +30:32:40.339  &  1.006  &  2   \\
--  &  B1  &  14:46:34.843  &  +30:32:19.925  &  0.579  &  3   \\
--  &  C1  &  14:46:33.538  &  +30:32:36.384  &  1.441  &  2   \\
--  &  D1  &  14:46:34.932  &  +30:33:13.133  &  ...  &   0  \\
--  &  E1  &  14:46:32.996  &  +30:33:10.256  &  ...  &   0  \\
--  &  F1  &  14:46:34.448  &  +30:32:49.265  &  ...  &   0  \\
--  &  G1  &  14:46:35.300  &  +30:33:02.009  &  ...  &   0  \\
--  &  H1  &  14:46:32.543  &  +30:32:53.584  &  ...  &   0  \\
--  &  gal  &  14:46:33.119  &  +30:33:18.365  &  0.4690  &  3  \\
--  &  gal  &  14:46:32.539  &  +30:32:51.297  &  0.4621  &  3  \\
--  &  gal  &  14:46:32.694  &  +30:31:50.069  &  0.4685  &  3  \\
--  &  gal  &  14:46:43.783  &  +30:33:45.838  &  0.4580  &  3  \\
SDSS J1456+5702  &  --  &   --  &  --  &  --  &  --  \\
--  &  A1  &  14:56:00.938  &  +57:02:35.127  &  0.8331  &  2   \\
--  &  A2  &  14:56:00.804  &  +57:02:12.193  &  0.8324  &  3   \\
--  &  B1  &  14:56:00.000  &  +57:02:05.642  &  ...  &  0   \\
--  &  B2  &  14:56:00.611  &  +57:02:27.382  &  ...  &  0   \\
--  &  C1  &  14:56:05.318  &  +57:02:05.821  &  1.141  &  2   \\
--  &  gal  &  14:56:13.136  &  +57:01:34.716  &  0.4883  &  3  \\
--  &  gal  &  14:55:59.784  &  +57:02:14.171  &  0.4865  &  3  \\
--  &  gal  &  14:56:08.707  &  +57:02:27.382  &  0.4864  &  3  \\
--  &  gal  &  14:56:05.339  &  +57:02:37.310  &  0.4952  &  3  \\
--  &  gal  &  14:56:01.329  &  +57:02:42.062  &  0.4933  &  2  \\
--  &  gal  &  14:56:05.559  &  +57:02:02.498  &  0.4733  &  3  \\
--  &  gal  &  14:56:00.368  &  +57:02:12.372  &  0.4816  &  2  \\
--  &  gal  &  14:55:59.114  &  +57:02:15.393  &  0.4728  &  2  \\
--  &  gal  &  14:55:57.051  &  +57:02:15.242  &  0.4765  &  3  \\
SDSS J1527+0652  &  --  &   --  &  --  &  --  &  --  \\
--  &  A1  &  15:27:48.950  &  +06:52:23.087  &  2.760  &  3   \\
--  &  A2  &  15:27:48.861  &  +06:52:23.520  &  2.760  &  3   \\
--  &  B1  &  15:27:46.647  &  +06:52:17.977  &  1.283  &  3   \\
--  &  gal  &  15:27:46.550  &  +06:51:57.778  &  0.3923  &  3  \\
--  &  gal  &  15:27:50.269  &  +06:51:20.769  &  0.3933  &  3  \\
--  &  gal  &  15:27:45.864  &  +06:52:56.098  &  0.3891  &  2  \\
--  &  gal  &  15:27:45.812  &  +06:52:33.273  &  0.3872  &  3  \\
--  &  gal  &  15:27:49.036  &  +06:50:53.806  &  0.3942  &  3  \\
--  &  gal  &  15:27:44.470  &  +06:52:22.150  &  0.3824  &  3  \\
--  &  gal  &  15:27:48.332  &  +06:51:04.966  &  0.3928  &  3  \\
--  &  gal  &  15:27:43.955  &  +06:52:44.337  &  0.3887  &  3  \\
--  &  gal  &  15:27:43.591  &  +06:53:10.498  &  0.3960  &  2  \\
--  &  gal  &  15:27:43.361  &  +06:53:35.158  &  0.3832  &  3  \\
--  &  gal  &  15:27:46.393  &  +06:51:29.447  &  0.3945  &  3  \\
--  &  gal  &  15:27:45.026  &  +06:51:35.422  &  0.3946  &  3  \\
SDSS J1531+3414  &  --  &   --  &  --  &  --  &  --  \\
--  &  A1  &  15:31:10.282  &  +34:14:14.640  &  1.097  &  2   \\
--  &  A2  &  15:31:09.849  &  +34:14:25.654  &  1.097  &  3   \\
--  &  A3  &  15:31:11.459  &  +34:14:34.182  &  1.097  &  3   \\
--  &  A4  &  15:31:11.693  &  +34:14:30.543  &  1.097  &  3   \\
--  &  A5  &  15:31:09.698  &  +34:14:55.605  &  1.097  &  3   \\
--  &  A6  &  15:31:07.559  &  +34:14:36.132  &  1.097  &  2   \\
--  &  B1  &  15:31:10.817  &  +34:14:38.865  &  1.300  &  3   \\
--  &  B2  &  15:31:08.287  &  +34:14:29.293  &  1.299  &  2   \\
--  &  C1  &  15:31:09.389  &  +34:14:08.804  &  1.0265  &  2   \\
--  &  gal  &  15:31:07.164  &  +34:13:12.499  &  0.3396  &  3  \\
--  &  gal  &  15:31:02.749  &  +34:14:33.235  &  0.3371  &  3  \\
--  &  gal  &  15:31:10.639  &  +34:15:20.270  &  0.3357  &  3  \\
--  &  gal  &  15:31:11.016  &  +34:14:28.785  &  0.3292  &  3  \\
--  &  gal  &  15:31:07.339  &  +34:16:41.637  &  0.3280  &  3  \\
--  &  gal  &  15:31:12.139  &  +34:14:05.467  &  0.3371  &  3  \\
--  &  gal  &  15:31:12.407  &  +34:13:55.524  &  0.3402  &  3  \\
--  &  gal  &  15:31:11.016  &  +34:14:28.799  &  0.3296  &  3  \\
SDSS J1621+0607  &  --  &   --  &  --  &  --  &  --  \\
--  &  A1  &  16:21:33.420  &  +06:07:14.865  &  4.1310  &  2   \\
--  &  A2  &  16:21:32.638  &  +06:07:05.470  &  4.1310  &  3   \\
--  &  B1  &  16:21:32.665  &  +06:07:18.789  &  1.1778  &  3   \\
--  &  C1  &  16:21:32.741  &  +06:07:30.994  &  ...  &  0   \\
--  &  gal  &  16:21:32.830  &  +06:07:11.275  &  0.3382  &  3  \\
--  &  gal  &  16:21:32.816  &  +06:07:14.120  &  0.3390  &  3  \\
--  &  gal  &  16:21:28.552  &  +06:06:53.276  &  0.3437  &  3  \\
--  &  gal  &  16:21:33.242  &  +06:07:26.968  &  0.3408  &  2  \\
--  &  gal  &  16:21:34.045  &  +06:07:23.408  &  0.3420  &  3  \\
--  &  gal  &  16:21:31.786  &  +06:07:44.468  &  0.3406  &  3  \\
--  &  gal  &  16:21:32.061  &  +06:07:48.572  &  0.3436  &  3  \\
--  &  gal  &  16:21:35.721  &  +06:06:29.192  &  0.3367  &  3  \\
--  &  gal  &  16:21:33.523  &  +06:07:15.163  &  0.3505  &  3  \\
--  &  gal  &  16:21:32.823  &  +06:07:25.604  &  0.3391  &  2  \\
--  &  gal  &  16:21:32.782  &  +06:07:28.737  &  0.3522  &  3  \\
SDSS J2111-0114  &  --  &   --  &  --  &  --  &  --  \\
--  &  A1  &  21:11:18.934  &  -01:14:31.427  &  2.858  &  2   \\
--  &  A2  &  21:11:20.280  &  -01:14:31.858  &  2.858  &  2   \\
--  &  B1  &  21:11:19.923  &  -01:13:56.398  &  1.476  &  3   \\
--  &  C1  &  21:11:19.395  &  -01:14:40.174  &  1.152  &  3   \\
--  &  gal  &  21:11:19.697  &  -01:13:30.728  &  0.6296  &  3  \\
--  &  gal  &  21:11:19.511  &  -01:13:53.704  &  0.6360  &  3  \\
--  &  gal  &  21:11:18.522  &  -01:12:51.172  &  0.6323  &  3  \\
--  &  gal  &  21:11:20.040  &  -01:14:00.219  &  0.6376  &  3  \\
--  &  gal  &  21:11:20.816  &  -01:14:41.411  &  0.6441  &  3  \\
--  &  gal  &  21:11:19.724  &  -01:15:26.598  &  0.6477  &  3  \\
SDSS J2238+1319  &  --  &   --  &  --  &  --  &  --  \\
--  &  A1  &  22:38:31.070  &  +13:19:46.946  &  0.724  &  3   \\
--  &  A2  &  22:38:31.448  &  +13:19:46.733  &  0.724  &  3   \\
--  &  A3  &  22:38:30.741  &  +13:19:58.468  &  0.724  &  3   \\
--  &  A4  &  22:38:30.933  &  +13:20:02.282  &  0.724  &  3   \\
--  &  A5  &  22:38:31.963  &  +13:19:58.286  &  0.725  &  3   \\
--  &  B1  &  22:38:30.603  &  +13:19:53.033  &  0.980  &  1   \\
--  &  B2  &  22:38:30.679  &  +13:19:56.013  &  0.980  &  3   \\
--  &  C1  &  22:38:31.441  &  +13:20:04.984  &  ...  &  0   \\
--  &  D1  &  22:38:31.777  &  +13:19:52.384  &  ...  &  0   \\
--  &  E1  &  22:38:31.771  &  +13:19:50.798  &  ...  &  0   \\
--  &  F1  &  22:38:30.720  &  +13:19:48.638  &  ...  &  0   \\
--  &  gal  &  22:38:31.214  &  +13:19:33.848  &  0.4089  &  3  \\
--  &  gal  &  22:38:38.788  &  +13:19:31.249  &  0.4112  &  3  \\
--  &  gal  &  22:38:30.754  &  +13:19:50.307  &  0.4087  &  3  \\
--  &  gal  &  22:38:30.178  &  +13:20:19.407  &  0.4094  &  3  \\
--  &  gal  &  22:38:30.596  &  +13:20:25.659  &  0.4118  &  3  \\
SDSS J2243-0935  &  --  &   --  &  --  &  --  &  --  \\
--  &  A1  &  22:43:25.181  &  -09:34:52.645  &  2.092  &  2   \\
--  &  A2  &  22:43:24.234  &  -09:35:10.034  &  2.093  &  3   \\
--  &  B1  &  22:43:23.300  &  -09:35:32.498  &  1.3202  &  3   \\
--  &  C1  &  22:43:24.440  &  -09:35:46.680  &  0.7403  &  2   \\
--  &  gal  &  22:43:23.540  &  -09:35:35.286  &  0.4413  &  3  \\
--  &  gal  &  22:43:23.629  &  -09:35:37.685  &  0.4466  &  3  \\
--  &  gal  &  22:43:24.391  &  -09:35:41.658  &  0.4423  &  3  \\
--  &  gal  &  22:43:26.390  &  -09:34:50.602  &  0.4536  &  2  \\
--  &  gal  &  22:43:26.280  &  -09:34:58.052  &  0.4447  &  3  \\
--  &  gal  &  22:43:19.626  &  -09:35:44.312  &  0.4457  &  3  \\
--  &  gal  &  22:43:25.298  &  -09:35:04.208  &  0.4560  &  3  \\
--  &  gal  &  22:43:29.775  &  -09:36:10.325  &  0.4455  &  3  \\
--  &  gal  &  22:43:26.699  &  -09:35:10.978  &  0.4493  &  3  \\
--  &  gal  &  22:43:24.282  &  -09:35:12.918  &  0.4513  &  3  \\
--  &  gal  &  22:43:20.464  &  -09:36:04.856  &  0.4464  &  3  \\
--  &  gal  &  22:43:32.988  &  -09:35:37.510  &  0.4461  &  2  \\
--  &  gal  &  22:43:32.968  &  -09:35:39.381  &  0.4462  &  3  \\
--  &  gal  &  22:43:20.718  &  -09:35:19.870  &  0.4499  &  3  \\
--  &  gal  &  22:43:23.210  &  -09:35:48.596  &  0.4492  &  3  \\
--  &  gal  &  22:43:19.303  &  -09:35:54.034  &  0.4437  &  3  \\
--  &  gal  &  22:43:24.131  &  -09:36:10.521  &  0.4543  &  3  \\
--  &  gal  &  22:43:27.124  &  -09:36:28.593  &  0.4398  &  3  \\
\enddata
\tablenotetext{a}{~Coordinates listed are J2000.0, with astrometry calibrated relative to the SDSS.}
\tablenotetext{b}{~See text for discussion of different redshift classifications.}
\end{deluxetable*}

\clearpage

\end{document}